\documentclass[11pt]{article}

\usepackage{booktabs}
\usepackage[english]{babel}
\usepackage{amsmath,amssymb,amsbsy,amstext, amsthm, simplewick}
\usepackage{hyperref}
\usepackage{graphicx}
\usepackage{amsfonts}
\usepackage{amssymb}
\usepackage{upgreek}
\usepackage{simplewick}
 \usepackage{exscale,relsize}

\usepackage[margin=1cm,labelfont={sf,bf,scriptsize},textfont={sf,scriptsize}]{caption}

\newcommand{\otoprule}{\midrule[\heavyrulewidth]}
\usepackage{colortbl}
\definecolor{lightgreen}{cmyk}{0.2, 0, 0.2, 0.2}
\definecolor{lightgray}{cmyk}{0.1,0.2,0,0.1}
\definecolor{lightgray2}{cmyk}{0.1,0.1,0,0.1}

\makeatletter
\newlength{\apb@width}
\newcommand{\autoparbox}[2][c]{\settowidth{\apb@width}{#2}\parbox[#1]{\apb@width}{#2}}

\makeatother

\numberwithin{equation}{section}

\def\beq{\begin{equation}}
\def\eeq{\end{equation}}

\def\bea{\begin{eqnarray}}
\def\eea{\end{eqnarray}}

\def\beq{\begin{equation}}
\def\eeq{\end{equation}}
\def\bea{\begin{eqnarray}}
\def\eea{\end{eqnarray}}

\def\Mp{M_{\rm pl}}

\def\H{{\cal H}}

\def\0{{\vec{0}}}
\def\k{{\vec{k}}}

\def\x{{\vec{x}}}

\def\cO{{\cal O}}
\def\k{{\bf k}}

\def\x{{\bf x}}

\def\cO{{\cal O}}

\DeclareRobustCommand{\SkipTocEntry}[4]{}

\def\Mp{M_{\rm pl}}

\DeclareSymbolFont{extraup}{U}{zavm}{m}{n}
\DeclareMathSymbol{\varheart}{\mathalpha}{extraup}{86}
\DeclareMathSymbol{\vardiamond}{\mathalpha}{extraup}{87}

\begin{document}

\begin{titlepage}

\setcounter{page}{1} \baselineskip=15.5pt \thispagestyle{empty}

\vbox{\baselineskip14pt
}
{~~~~~~~~~~~~~~~~~~~~~~~~~~~~~~~~~~~~
~~~~~~~~~~~~~~~~~~~~~~~~~~~~~~~~~~
~~~~~~~~~~~ \footnotesize{SLAC-PUB-15334, SU/ITP-12/42}} \date{}

\bigskip\

\vspace{2cm}
\begin{center}
{\fontsize{19}{36}\selectfont  \sc Anomalous dimensions and non-gaussianity}
\end{center}

\vspace{0.3cm}

\begin{center}
{\fontsize{13}{30}\selectfont  Daniel Green$^{\blacklozenge, \spadesuit}$, Matthew Lewandowski$^ \blacklozenge$, Leonardo Senatore$^{\blacklozenge, \varheart, \spadesuit, \bigstar}$,\\  Eva Silverstein$^{\blacklozenge, \varheart,\spadesuit}$, and Matias Zaldarriaga$^\clubsuit$}
\end{center}


\begin{center}
\vskip 8pt
\textsl{$^ \blacklozenge$
Stanford Institute for Theoretical Physics, Stanford University, Stanford, CA 94306, USA}

\vskip 7pt
\textsl{$^\varheart$ SLAC National Accelerator Laboratory, 2575 Sand Hill, Menlo Park, CA 94025}

\vskip 7pt
\textsl{$^\spadesuit$ Kavli Institute for Particle Astrophysics and Cosmology, Stanford, CA 94025, USA}

\vskip 7pt
\textsl{$^\bigstar$ CERN, Theory Division, 1211 Geneva 23, Switzerland}

\vskip 7pt
\textsl{$^\clubsuit$ School of Natural Sciences,
 Institute for Advanced Study,
Princeton, NJ 08540, USA}

\end{center}

\vspace{1.2cm}
\hrule \vspace{0.3cm}
{ \noindent \textbf{Abstract} \\[0.2cm]
\noindent 
We analyze the signatures of inflationary models that are coupled to strongly interacting field theories, a basic class of multifield models also motivated by their role in providing dynamically small scales.  
Near the squeezed limit of the bispectrum, we find a simple scaling behavior determined by operator dimensions,
which are constrained by the appropriate unitarity bounds.
Specifically, we analyze two simple and calculable classes of examples: conformal field theories (CFTs), and large-N CFTs deformed by relevant time-dependent double-trace operators.     Together these two classes of examples exhibit a wide range of scalings and shapes of the bispectrum, including nearly equilateral, orthogonal and local non-Gaussianity in different regimes.  Along the way, we compare and contrast the shape and amplitude with previous results on weakly coupled fields coupled to inflation.  This signature provides a precision test for strongly coupled sectors coupled to inflation via irrelevant operators suppressed by a high mass scale up to $\sim 10^3$ times the inflationary Hubble scale. }

 \vspace{0.3cm}
 \hrule

\vspace{0.6cm}
\end{titlepage}

\tableofcontents

\newpage
\section{Introduction and summary}

In inflationary cosmology, the quantum fluctuations of the inflaton, and possibly other fields, get imprinted on the power spectrum and higher point correlations accessible in the CMB and large-scale structure.  This basic idea goes back to the origins of the subject, and has been explored in many illustrative examples.  Recently there has been progress toward a more systematic understanding of the observables and their implications, with input both from bottom up effective field theory and from top down UV complete mechanisms for inflation.

In this paper, we mix in a basic class of strongly-coupled fields -- field theories which are conformal at the Hubble scale and their time-dependent deformations -- and compute their contribution to the perturbation spectrum and non-Gaussianity.    We find characteristic scaling behavior depending on the anomalous dimensions of operators, and compute in detail the shape and amplitude of the non-Gaussianity.  The results depend on some basic properties of quantum field theory, such as unitarity bounds and crossing relations.  

Aside from theoretical interest, this analysis is motivated by several considerations.  First, it is a canonical and calculable regime of field theory which belongs in a systematic analysis of multifield signatures and their significance, particularly given the observational handles on non-Gaussianity.  Relatedly, as we will see, these strongly coupled sectors exhibit an interesting partial degeneracy with weakly interacting fields in their predictions for the simplest practical observables.  Finally, from the top down, strongly coupled sectors play a useful role in model-building, producing naturally small scales in the effective action for the inflaton, and fields from this sector may participate in the perturbations.  More generally, we can use forthcoming non-Gaussianity data (whether a constraint or detection) to provide a precision test of additional field theory sectors which couple to inflation via higher dimension operators.   
  
To put this in context, one basic question probed by observations is the number of fields which participate in generating the correlation functions of the curvature perturbations.  This is a different question from the question of how many fields are involved in the underlying mechanism producing the inflationary background.  One can separate multi-field theories into two classes, depending on whether the additional fields themselves acquire scale invariant perturbations that affect the curvature or the isocurvature fluctuations.  In the case we will be interested in here, the additional fields instead affect the density perturbations through their couplings to the inflaton. Though not directly observable, these additional fields cannot usefully be integrated out, as doing so would lead to a non-local Lagrangian.  Progress toward a general treatment of both types of theories can be found in  \cite{Senatore:2010wk}\cite{LopezNacir:2011kk}, 
where the effects of additional sectors are packaged in terms of their correlation functions.          

A useful general theorem known as the consistency condition establishes that a single-field theory of perturbations cannot generate non-Gaussianity in the `squeezed' configuration, with one mode much longer than the others \cite{Juan}.  
Conversely, it has been long known that non-Gaussianity peaked in the squeezed configuration can arise in the presence of additional light fields \cite{fnLmulti}.  For massless weakly interacting fields coupled to the inflaton, the three point function of scalar perturbations behaves as
\beq\label{localsqueezed}
\langle \zeta_{k_1}\zeta_{k_2}\zeta_{k_3}\rangle \xrightarrow{k_1\ll \, k_2 \simeq k_3}  \frac{f_{NL}\Delta_\zeta^4}{k_1^3 k_2^3} \ (2 \pi)^3\delta^{(3)}(\k_1+\k_2+\k_3)
\eeq
where $\Delta_\zeta\approx 10^{-5}$ is the amplitude of the scalar perturbation.  
Moreover, interesting power-law deviations from this shape of non-Gaussianity arise for fields of nonzero mass $m$   \cite{QSFI}
\bea
\langle \zeta_{k_1}\zeta_{k_2}\zeta_{k_3}\rangle &=& B(k_1,k_2,k_3) \ (2\pi)^3 \delta^{(3)}(\k_1+\k_2+\k_3) \\
 &\xrightarrow{k_1\ll \, k_2 \simeq k_3}&  \frac{ f_{NL}\Delta_\zeta^4}{k_1^3 k_2^3}\times \left(\frac{k_1}{k_2}\right)^{3/2-\sqrt{9/4-m^2/H^2}} \ (2\pi)^3 \delta^{(3)}(\k_1+\k_2+\k_3)\label{massive}
\eea
This scenario, known as quasi single-field inflation, is a feature of weakly coupled theories with supersymmetry broken at the Hubble scale \cite{Dans}.      

In the present work, we will analyze two simple cases where the additional fields affecting the density perturbations are strongly interacting, using conformal symmetry and related methods to control the calculations.  For our first class of examples, we will study inflation coupled to a conformally coupled CFT, for which the conformal dimension $\Delta$ of an operator $\cO$ plays the role of the mass-dependent exponent in (\ref{massive}).  For our second class of examples we will consider a simple time-dependent flow away from a large-N CFT, which gives unitary theories realizing a larger range of exponents, including local $f_{\rm NL}$.   

For our first class of examples, a conformally coupled CFT linearly mixes with inflaton perturbations \footnote{Note that our CFT lives in the four-dimensional spacetime, and is not to be confused with conjectural lower-dimensional holographic duals for de Sitter.} , giving rise to a bispectrum of the form
\bea\label{OpsNG}
B(k_1,k_2,k_3) &\xrightarrow{k_1\ll \, k_2 \simeq k_3}&  \frac{ f_{NL}\Delta_\zeta^4}{k_1^3 k_2^3}\times \left(\frac{k_1}{k_2}\right)^\Delta ~~~~ (\Delta\le 2) \\
 &\xrightarrow{k_1\ll \, k_2 \simeq k_3}&  \frac{ f_{NL}\Delta_\zeta^4}{k_1^3 k_2^3}\times \left(\frac{k_1}{k_2}\right)^2 ~~~~ (\Delta\ge 2) \ .  \nonumber
\eea
For this case, the standard unitarity bound implies $\Delta\ge 1$.  As a result, the shape is peaked at equilateral/flattened triangles in momentum space, with the scaling behavior (\ref{OpsNG}) as one approaches the squeezed limit determined by the dimension of the most relevant operator that couples in.  Moreoever, we will find a range of exponents arising from dimensions $3/2\leq\Delta\leq 2$ which cannot be obtained in quasi-single field models, producing a scale-dependent bias which could distinguish them.  For the range of exponents where the two are degenerate, it is intriguing that our intrinsically gapless CFT fields behave like massive weakly coupled fields with respect to the squeezed limit; the origin of this effect is the redshifting of conformally rescaled correlators of our conformally coupled operators. The conformal coupling to curvature which goes into this analysis is a special choice, and we expect a wider range of behaviors in the presence of more general curvature couplings.    

Indeed, our second class of examples will give rise to the same scaling\footnote{up to logarithmic factors.} as (\ref{OpsNG}) while allowing for $\Delta <1$.   These examples make use of the fact that time-dependent couplings in quantum field theory can strongly affect infrared physics and shift unitarity bounds, as studied recently in \cite{ubounds}.  This is important in the present context since time dependent couplings can arise very easily via couplings of the rolling inflaton field to other sectors such as a CFT.  In the examples in \cite{ubounds},  time dependent couplings can introduce flows between a unitary CFT containing an operator ${\cal O}$ of dimension $\Delta_+$  and an infrared theory with two-point correlators falling off like $1/distance^{2\Delta_-}$ times powers of the time-dependent coupling, where $\Delta_-=4-\Delta_+$.  In particular a theory with a marginal  scalar operator with $\Delta_+\approx 4$  can flow in this way to a theory with $\Delta_-\ll 1$, giving nearly local non-Gaussianity, not suppressed in the squeezed limit by any additional powers of $k_1/k_2$.    

Finally, let us discuss the amplitude $f_{NL}$ of the bispectrum.  As we will see, this can be substantial, and will give us sensitivity to higher dimension couplings of the inflationary sector to other fields.  
To give a rough illustration of this last point, consider, for example, a dimension $\Delta=2$ operator $\cO_2$ in a CFT, coupled to the inflaton $\phi$ via the dimension six operator
\beq\label{dimsix}
\int d^4x \sqrt{-g}\frac{(\partial\phi)^2\cO_2}{M_*^2} \ .
\eeq 
Let us evaluate one factor of $\partial\phi$ on the backround rolling scalar field, which in slow-roll inflation given in terms of the inflationary Hubble scale $H$ by $\dot\phi_0\sim H^2/\Delta_\zeta\sim 10^5 H^2$.  This gives us a linear mixing $\int \Delta_\zeta^{-1} (H/M_*)^2\delta \dot\phi \, \cO$ between the canonical perturbation $\delta\phi$ and $\cO$.\footnote{For the purposes of finding a simple estimate of our sensitivity to high dimension couplings, we will tune away the relevant perturbation of the CFT that we get from (\ref{dimsix}) if we evaluate both factors of $\partial\phi$ on the rolling inflaton background.  In the detailed examples to be discussed in the main body of this paper and appendix A, we will find similar results without such tuning.}      The CFT three-point function combined with three of these mixing interactions generates a contribution to the three-point function of the inflationary perturbations.  This leads to an amplitude $f_{NL}$ behaving parametrically like
\beq\label{CFTfNLscaling}
f_{NL}\sim C\Delta_\zeta^{-4} f(2) \left(\frac{H}{M_*}\right)^6  
\eeq    
where $C$ is the amplitude of the CFT three-point function (\ref{equ:cft2}), and the function $f(\Delta)$, plotted below in Fig. 6, gives a substantial numerical factor.\footnote{Below we will describe our normalization conventions which go into this.}  Since current measurements are projected to be sensitive to $f_{NL}\lesssim 10$, they provide a precision test of this higher dimension coupling up to a value of $M_*\sim 10^3 H$.  This high mass scale (relative to Hubble) is generally below the Planck scale ($\Mp$), but can exceed the GUT scale in simple examples such as chaotic inflation.  In string-theoretic ultraviolet completions of inflation, higher dimension operators suppressed by a scale $M_*\ll \Mp$ of order the Kaluza-Klein or string scale may be probed.\footnote{In several of the UV completions of inflation explored extensively in string theory, reviewed for example in \cite{inflthroats}, strongly coupled sectors play a useful role in producing dynamically small scales  \cite{GKPKKLT}.  Operators suppressed by the Kaluza-Klein scale are ubiquitous there \cite{tunneling}.  (See  \cite{QCDmonodromy}\ for some more recent examples in which interacting field theory plays a role in the inflationary mechanism.)}  That is, while an observation of non-Gaussianity may be explained by contributions from additional fields (weakly or strongly coupled),  a null result would conversely provide a precision constraint on very high-energy physics.  
Of course sensitivity to high energy physics arises already in single field inflation (see e.g. \cite{DBI,generalsingle,EFT,Senatore:2009gt}). A feature of the present case is that observations will constrain hidden sectors of additional fields coupled to inflation via higher dimension operators such as (\ref{dimsix}).  Similar constraints arise in other models involving a linear mixing with the inflaton (e.g.~\cite{QSFI, turning}).

Both of our main examples are just calculable examples of a wider point:  additional fields active during inflation may include strongly coupled sectors with characteristic signatures.  It will be interesting to analyze this more generally and systematically, and also to incorporate the particular couplings arising in complete models of inflation which involve couplings to strong dynamics to obtain their specific multifield signatures.

\section{Conformally Coupled Examples}\label{sec:conformal}

In this section, we will consider strongly coupled theories which behave like a CFT near the inflationary Hubble scale.   
In flat spacetime, CFTs are one of the best studied classes of interacting field theories.  Due to the high degree of symmetry, much is known about the spectrum of operators and their correlations functions.  Because de Sitter space is conformally flat, a CFT can be coupled to gravity such that correlation functions in de Sitter space preserve the flat space results up to an overall rescaling.  This choice is related to the choice of curvature couplings involving operators in the CFT (or equivalently to the choice of improvement terms for the stress tensor that we couple to gravity).  In this section, we will assume these couplings are chosen to preserve conformal invariance.  

\subsection{Setup}

We are interested in the possibility that a CFT is weakly coupled to the inflaton and its perturbation.  As a result, the CFT will influence the correlation functions of the curvature perturbation we observe at late times.  For this purpose, it is not necessary to specify the underlying dynamics leading to inflation; instead it is most convenient to work directly in terms of the perturbations $\pi(t,\vec x)$ as in the effective field theory treatment developed in \cite{EFT}.  Indeed, the effects we wil compute could accompany a wide variety of underlying inflationary mechanisms.     

In the absence of inflation, a CFT is described by a list of local, primary operators ${\cal O}_i^{(j,\tilde j), \Delta_i}(\x,t)$  and their correlation functions, where $(j,\tilde j) \in (\frac{\mathbb Z}{2} , \frac{\mathbb Z}{2})$ is the spin and $\Delta_i$ is the dimension of the $i$-th operator.  For simplicity we will focus on a single scalar operator $\cO(\x,t)$ with dimension $\Delta$.  In flat space, the two- and three-point functions of $\cO(\x,t)$ are fixed up to a constant.  Specifically, the correlation functions in Euclidean signature take the form
\bea\label{equ:cft2}
\langle \cO(\x, t) \cO(0) \rangle& =& \frac{1}{|x^2 + t^2|^{ \Delta} }\ , \\ \langle \cO(\x_1,t_1) \cO(\x_2, t_2) \cO(\x_3,t_3) \rangle &=& \frac{ C}{|x_{12}^2 + t^2_{12}|^{\tfrac{\Delta}{2}} |x_{23}^2 + t^2_{23}|^{\tfrac{\Delta}{2}} |x_{31}^2 + t^2_{31}|^{\tfrac{\Delta}{2}} } \label{equ:cft3}\ ,
\eea
where $\x_{ij} = \x_i - \x_j$, $t_{ij} =t_i - t_j$ and $C$ is a constant.  The normalization of the two-point function is a convention~\footnote{In the limit $\Delta=1$, the CFT becomes free. In this case, the convention for the two-point function differs from the one of a free scalar field a factor of $4\pi^2$. This explains why our dimensionless functions have sometimes large numerical values.}.   

The metric of de Sitter space is given by
\beq
ds^2 = a(\tau)^2 (-d\tau^2 + d\x^2) \ .
\eeq 
where $a(\tau)=-1/(H\tau)$.
This metric is conformally flat in terms of the conformal time $\tau$.  Therefore, we get the CFT correlation functions in de Sitter space from the flat space result by replacing $t \to \tau$ and a Weyl transformation of $\cO(x,t) \to a(\tau)^{-\Delta} \cO(\x, \tau)$.
In particular, the two and three point functions are given by (in Euclidean time):
\bea\label{dStwo}
\langle {\cal O}(\tau, \x){\cal O}(\tau',\x') \rangle &=& \frac{a(i\tau)^{-\Delta} a(i\tau')^{-\Delta}}{\left[(\tau-\tau')^2+( x- x')^2\right]^\Delta} \\ \label{dSthree}
\langle {\cal O}(\x_1,\tau_1){\cal O}( \x_2, \tau_2){\cal O}( \x_3,\tau_3)\rangle &=& \frac{C\; a(i\tau_1)^{-\Delta}a(i\tau_2)^{-\Delta}a(i\tau_3)^{-\Delta}  }{|x_{12}^2+\tau_{12}^2|^{\Delta/2} | x_{23}^2+\tau_{23}^2|^{\Delta/2} | x_{31}^2+\tau_{31}^2|^{\Delta/2} }
\eea
In particular, these correlation functions redshift at late times much like those of a weakly interacting massive field.  These strongly coupled correlators are different from those that arise in the weakly coupled case \cite{QSFI}\cite{Dans}; the position space two point function for massive fields is a nontrivial hypergeometric function, distinct from the simpler function (\ref{dStwo}) except for the special case of a conformally coupled free scalar.  

We will now couple the inflationary perturbation $\pi$ to this CFT via the interaction Hamiltonian
\beq\label{equ:doublemix0}
 {\cal H}_{\rm int} =\tfrac{1}{2}  \mu^{2 - \Delta} \Mp 
|\dot H|^{1/2} ( 2 \dot \pi - \partial_\mu \pi \partial^\mu \pi) \cO + \tfrac{1}{4} \Mp^2 |\dot H| \tilde \mu^{- \Delta} ( - 2 \dot \pi + \partial_\mu \pi \partial^\mu \pi)^2 \cO  \ .
 \eeq
Here and elsewhere, $\dot f=- H\tau \tfrac{df}{d\tau}$ is a derivative with respect to FRW time $t=-\log(-H\tau)/H$.   The scalar $\pi\sim \delta t$ is not canonically normalized; it is related to the canonically normalized\footnote{Here we are assuming unit speed of sound, $c_s \sim 1$.  For general sound speed, the canonically normalized field is given by $\pi=c_s \pi_c/\sqrt{2\Mp^2\dot H}$.} perturbation $\pi_c$ via $\pi=\pi_c/\sqrt{2\Mp^2\dot H}$.  It is also related to the conventionally normalized scalar perturbation $\zeta$ via $\zeta = - H \pi$ (at linear order).  We also note here that at freezeout, the corresponding field amplitudes which will enter into the calculations below are $\zeta\sim \Delta_\zeta\sim 10^{-5}$ and $\pi_c\sim H$.  In the whole paper we will neglect the mixing with gravity, as it will give subleading corrections.

In a slow-roll model like (\ref{dimsix}), the parameters $\mu$ and $\tilde \mu$ may be related to $\dot \phi$ and some higher scale, $M_*$ (e.g. $\mu = \dot \phi / M_*$ with $M_*^2 \gg \dot \phi$ when $\Delta =1$).  However, there may be many other UV completions that also give rise to (\ref{equ:doublemix0}) where these scales have different origins.  For this reason, we will work directly with $\mu$ and $\tilde \mu$ throughout, as they are the parameters relevant to the phenomenology.  

When $\Delta < 2$ the leading contribution from the first term is a relevant deformation and therefore is perturbative when $\mu \ll H$.  On the other hand, the second term is irrelevant for all dimensions consistent with unitarity and is therefore perturbative the $\tilde \mu \gg H$.  A priori, the second term may or may not contribute significantly to the bispectrum.  We will therefore consider the two cases separately in section \ref{sec:boo} and \ref{sec:booo}.  For the special case of $\Delta =2$, we should replace $(\mu / H)^{2-\Delta}$ with a dimensionless coupling $\lambda$: $\lambda=\lim_{\Delta\to2} (\mu/H)^{2-\Delta}$, with $\log(\mu/H)\to (\lambda-1)/(\Delta-2)$.  This replacement should be unambiguous so we will not do it explicitly.  

In Appendix A, we analyze the radiative stability of this setup.  One result of that analysis is that under appropriate conditions the term $\sim\int m^{4-\Delta}\cO$ generated by $\pi$ loops satisfies $m\ll H$, meaning that even for relevant operators ($\Delta<4$) we do not generate a flow away from the CFT over the scales of interest.  This analysis is self-contained up to a scale $\Lambda$ which can be $\gg H$, leading to precision tests of higher dimension operators as anticipated in the introduction.

\subsection{Calculating $in$-$in$ Correlators in Euclidean Signature}

Throughout the paper, we will be interested in calculating $in$-$in$ correlations functions of $\zeta$ evaluated at equal times.  These can be computed perturbatively, using the interaction picture fields, stating from~\cite{Weinberg:2005vy}, 
\beq\label{equ:inin}
\langle \bar T \exp[i \int^{\tau_0}_{-\infty(1+i\epsilon)}  H_{\rm int}(\tau)a(\tau)d\tau ]\Big( \zeta_{\rm int}(\k_1, \tau_0) .. \zeta_{\rm int}(\k_n, \tau_0)  \Big)  T \exp[ -i \int^{\tau_0}_{-\infty(1-i\epsilon)}  H_{\rm int}(\tau) a(\tau) d\tau ] \rangle \ ,
\eeq
where we have assumed the Bunch-Davies vacuum and $H_{\rm int} (\tau)= \int d^3 x \, a^3(\tau) \H_{\rm int}(\tau,x)$.  In order to simplify these calculations, we will follow the strategy suggested in \cite{Behbahani:2012be}.  The basic idea is that,  after rotating the conformal time integrals via $\tau \to \pm i  \tau_E + \tau_0$, the $in$-$in$ correlation becomes a (Euclidean) anti-time-ordered correlation function.  

\begin{figure}[h!]
   \centering
       \includegraphics[scale =0.35]{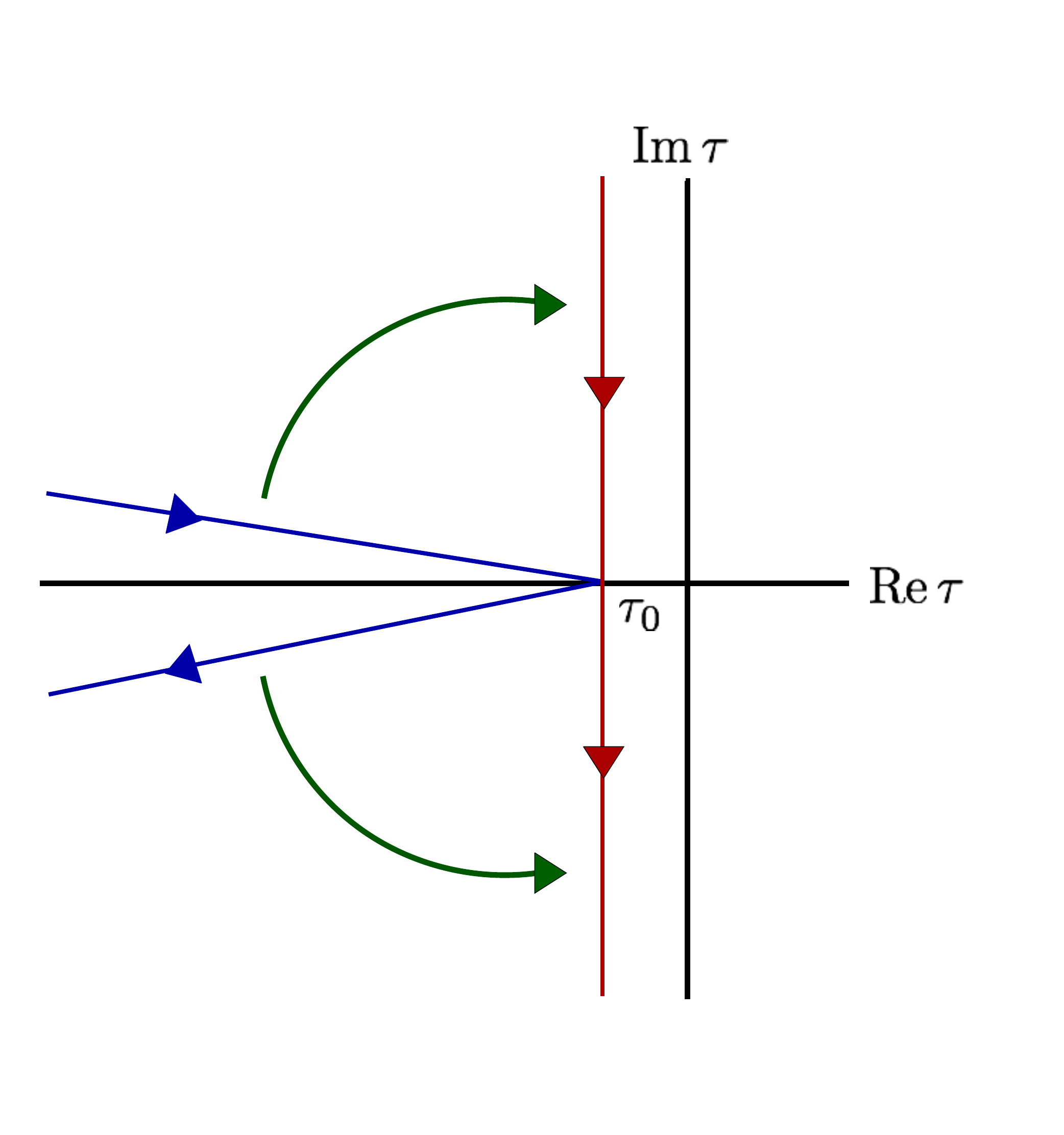}
   \caption{The analytic continuation of the contour in conformal time ($\tau$) from Lorentzian signature (blue) to Euclidean signature (red).  We calculate the correlation functions for operators at $\tau_0 < 0$.  Our expressions involve branch cuts only when ${\rm Re} \tau > 0$, which ensures this continuation is well defined. }
  \label{fig:contour}
\end{figure}

Starting from the $i \epsilon$ prescription in (\ref{equ:inin}), which projects onto the interacting vacuum, the time and anti-time ordered exponentials should be rotated to opposite values of the Euclidean time.  After doing this, we obtain
\beq\label{npoint}
\langle \bar T \exp[i \int^{\tau_0}_{-i\infty +\tau_0}  H_{\rm int}(\tau) a(\tau) d\tau ]\Big( \zeta_{\rm int}(\k_1, \tau_0) .. \zeta_{\rm int}(\k_n, \tau_0)  \Big)  T \exp[ -i \int^{\tau_0}_{i \infty +\tau_0}  H_{\rm int}(\tau) a(\tau) d\tau ] \rangle \ .
\eeq
This we recognize as simply the anti-time ordered correlation function in Euclidean time
\bea
&&\langle \bar T \Big( \zeta_{\rm int}(\k_1, \tau_0) .. \zeta_{\rm int}(\k_n, \tau_0) \exp[ i \int^{ i \infty+\tau_0}_{-i \infty +\tau_0}  H_{\rm int}(\tau) a(\tau) d\tau ] \Big)   \rangle \\
&& = \langle \bar T \Big( \zeta_{\rm int}(\k_1, \tau_0) .. \zeta_{\rm int}(\k_n, \tau_0)  \exp[ - \int^{ \infty}_{- \infty }  H_{\rm int}(  i\tau_E + \tau_0) a(i \tau_E + \tau_0)  d\tau_E ] \Big)   \rangle \ .
\eea
Provided we use the anti-time ordered, Euclidean Green's function which we will compute momentarily, the operator ordering is automatic.  This analytic continuation is illustrated in figure \ref{fig:contour}.

The anti-time-ordered Green's functions for any two operators can be written as 
\beq
\langle {\bar T} \cO_1(\tau) \cO_2(\tau') \rangle\equiv \theta(\tau'_E- \tau_E )\langle  \cO_1(\tau) \cO_2(\tau') \rangle+\theta( \tau_E-\tau'_E )\langle  \cO_2(\tau') \cO_1 (\tau) \rangle \ ,
\eeq
where $\theta(x)$ is the Heaviside step function.   Following the usual quantization of a scalar in de Sitter space, we write the operator $\hat \pi_{\k} = \pi_{k} (\tau) \hat a_\k^{\dagger} + \pi_{k}^*(\tau) \hat a_\k$ with $[\hat a_\k , \hat a_{\k'}^{\dagger} ] = (2\pi)^3 \delta(\k + \k')$ and
\beq
\pi_{k} (\tau) = \frac{H}{2 \Mp |\dot H|^{1/2}} \frac{(1- i k \tau)}{k^{3/2}} e^{i k \tau} \ ; ~~~~ \pi_{k}^* (\tau) = \frac{H}{2 \Mp |\dot H|^{1/2}} \frac{(1+ i k \tau)}{k^{3/2}} e^{-i k \tau} \ .
\eeq
 (As usual, we conjugate $\pi$ to obtain $\pi^*$ in our original Lorentzian signature calculation, before deforming our contour to lie along $\tau_E$.)
 Notice that external factors of $\pi_\k(\tau_0)$ only have nonzero contractions with factors of $\dot\pi_c$ in $\cal{H}_{\rm{int}}$, so that we will only need
\beq
 \langle \bar T\Big(\dot \pi_c(i \tau_E+ \tau_0, \k_1) \pi_c( \tau_0, \k_2 ) \Big) \rangle  = - \frac{H^3}{2 k_1}(1+ i k_1 \tau_0 \frac{\tau_E}{|\tau_E| } )(-i \tau_E- \tau_0)^2e^{- k_1|\tau_E | }  (2\pi)^3 \delta(\k_1 + \k_2)\ . 
\eeq
The appearance of the absolute values in the above expression does not obstruct the analytic continuation we performed in writing (\ref{npoint}); the only non-analytic behavior appears at the location of the operator insertions, $\tau = \tau_0$, which is fixed in our continuation.  Alternatively, one can choose the domains of integration in (\ref{equ:inin}) to be manifestly time-ordered, which ensures the integrands are analytic in $\tau$.  

As we discussed in the previous section, the CFT correlation functions in Euclidean signature may be be taken to be anti-time ordered.   Therefore, when evaluating correlation functions we can use equations (\ref{dStwo}) and (\ref{dSthree}) with $\tau_i = i \tau_{i, E} + \tau_0$ (we will drop the $_E$ in remainder of this section).  Since we are interested in behavior the correlation functions at late times, we will take $\tau_0 \to 0$ at the end of all our calculations.

\subsection{Corrections to the Power Spectrum}

With the coupling ${\cal H}_{\rm int} \supset \tfrac{1}{\sqrt{2}} \mu^{2-\Delta} \dot \pi_c \cO$, we expect a correction to the power spectrum  ${\cal P}_\zeta\sim \Delta_\zeta^2/k^3$ at order $(\tfrac{\mu}{H})^{4-2 \Delta}$.  The action does not depend explicitly on time and therefore we expect this correction to be scale invariant on general grounds.  In this subsection, we will confirm this intuition with an explicit calculation.

\begin{figure}[h!]
   \centering
       \includegraphics[scale =0.9]{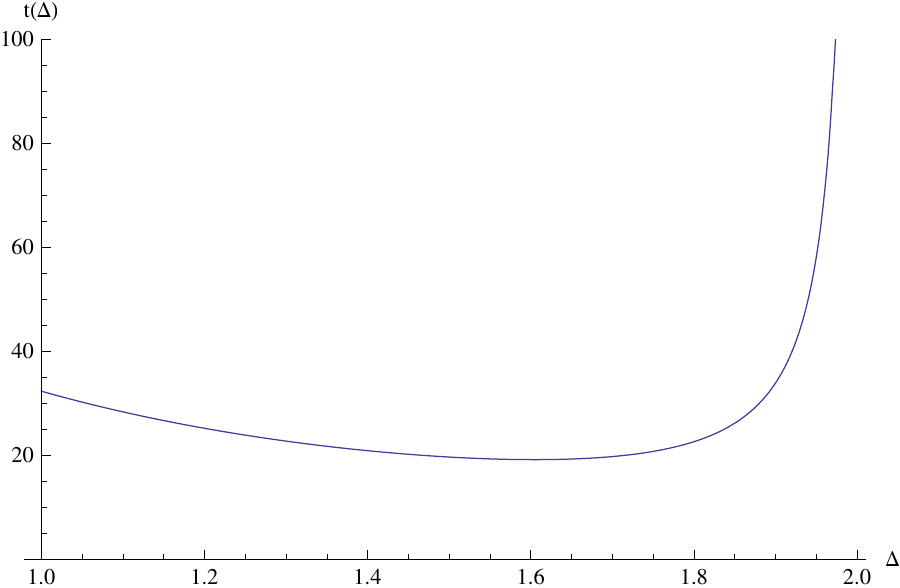}
   \caption{Numerically computed $t(\Delta)$.  }
  \label{fig:power1}
\end{figure}

The correction to the power spectrum arises from
\bea\label{eq:power_correction}
&&\delta {\cal P}_\zeta =\Mp^2 |\dot H| \mu^{4-2 \Delta} \langle \zeta_\k(\tau_0) \zeta_{-\k}(\tau_0)  \int  d \tau_1 d \tau_2 \;a(\tau_0+i \tau_1)^4 a(\tau_0+i \tau_2)^4 \\ \nonumber
&&\qquad\times\    \ \dot\pi_\k(\tau_0+i \tau_1) \dot\pi_{-\k}(\tau_0+i \tau_2) \rangle\ \langle \cO_{-\k} (\tau_0+i \tau_1)\cO_{-\k} (\tau_0+i \tau_2) \rangle' \ , 
\eea
where the prime on $\langle \cO \cO\rangle'$ indicates that we drop the $(2\pi)^3$ times a delta function in momentum.  We will use the identity
\beq\label{eq:real-2-point}
\frac{1}{(x^2)^{\Delta}} =\frac{ (2\pi)^2}{4^{\Delta-1}} \frac{\Gamma(2-\Delta)}{ \Gamma(\Delta)}\int \frac{d^4k}{(2\pi)^4} e^{i k \cdot x} (k^2)^{\Delta-2} \ ,
\eeq
where $x$ is a 4-vector.  The two-point function of $\cO$ can be written in momentum space as
\bea
&&\langle \cO_\k (\tau_0+i \tau_1)  \cO_{-\k} (\tau_0+i \tau_2) \rangle' = \\ \nonumber
&&\quad\quad \frac{ (2\pi)^2}{4^{\Delta-1}} \frac{\Gamma(2-\Delta) }{ \Gamma(\Delta)}a(\tau_0+i \tau_1)^{-\Delta} a(\tau_0+i \tau_2)^{-\Delta} \int\frac{ d\omega}{2\pi} e^{ i \omega \tau_{12} } (k^2 +\omega^2)^{\Delta - 2} \ . \nonumber 
\eea
Plugging back in (\ref{eq:power_correction}) the two integrals in $\tau_1$ and $\tau_2$ can be done analytically, and we are left with the integral in $\omega$ to be done numerically. The correction to the power spectrum is then given by
\bea
\delta {\cal P}_\zeta &\equiv& {\cal P}_\zeta(k) \Big( \frac{\mu}{H} \Big)^{4-2 \Delta} t(\Delta)
\eea
where 
\bea
&&t(\Delta)=-\frac{\pi ^2 4^{1- \Delta} \Gamma (2-\Delta) }{\Gamma (\Delta) }  e^{-i (2 \tilde\tau_0+\pi  \Delta)} \int d\tilde\omega\;\frac{1}{\left(\tilde\omega^2+1\right)^{2+\Delta} }\ \times\\ \nonumber
&&\left[(\tilde\tau_0+i) (\tilde\omega-i) e^{i (2
   \tilde\tau_0+\pi  \Delta)} (1-i  \tilde\omega)^\Delta \Gamma (\Delta-1,-\tilde\tau_0
   (\tilde\omega-i))\right.\\ \nonumber
  &&\qquad\qquad\qquad\qquad\qquad \left.+(\tilde\tau_0-i) (\tilde\omega+i) (1+i  \tilde\omega)^\Delta
   \Gamma (\Delta-1,-\tilde\tau_0 (\tilde\omega+i))\right]\ \times \\ \nonumber
   &&\left[(\tilde\tau_0-i)
   (\tilde\omega-i) (1-i  \tilde\omega)^\Delta \Gamma (\Delta-1,\tilde\tau_0 (\tilde\omega-i))\right.
   \\ \nonumber
   &&\qquad\qquad\qquad\qquad\qquad\left.+(\tilde\tau_0+i) (\tilde\omega+i) e^{i (2 \tilde\tau_0+\pi  \Delta)} (1+i 
   \tilde\omega)^\Delta \Gamma (\Delta-1,\tilde\tau_0 (\tilde\omega+i))\right]\ ,
   \eea
where $\Gamma[s,x]  \equiv \int_x^{\infty}  t^{s-1} e^{- t}  d t$ is the upper incomplete gamma function, and 
where we take $\tilde\tau_0=\tau_0/k \to 0$ at the end of the calculation.  In the limit $\tilde\tau_0 \to 0$, $t(\Delta)$ is independent of $k$ and therefore the power spectrum remains scale invariant.  The function $t(\Delta)$ is plotted in figure \ref{fig:power1}. Notice that its numerical value is quite large, though the correction to the power spectrum is safely much smaller than one for a large range of values for $\mu$. The large value is partly related to our conventions in equation \ref{equ:cft2}, which differ from free field conventions by a factor of $4 \pi^2 \simeq 39.5 $.  Further, notice the divergence as $\Delta\to 2$. This is due to the necessity of a divergent counter-term for the two point function for $\Delta\geq 2$. In the language of the EFT of Inflation, the unitary gauge operator that provides the counter-term is $(\delta g^{00})^2$~\cite{Senatore:2009cf}. Upon re-insertion of $\pi$, this terms contains indeed the quadratic term $\dot\pi^2$. We see that a speed of sound different from unity is generated.  We discuss about radiative corrections and renormalization in larger detail in Appendix~\ref{app:tuning}. We conclude that, upon renormalization, the contribution is small even for $\Delta>2$.

\subsection{Bispectrum from $\langle\cO\cO\rangle$} \label{sec:boo}
Let us begin to explore signatures in the bispectrum. The simplest case to compute is when the bispectrum in $\pi$ (and so in $\zeta$) is induce by the power spectrum (two point function) of $\cO$'s.  This arises from a combination of two types of vertices from the following interaction Hamiltonian density
\beq\label{equ:doublemix0a}
 {\cal H}_{\rm int} =\tfrac{1}{2}  \mu^{2 - \Delta} \Mp 
|\dot H|^{1/2} ( 2 \dot \pi - \partial_\mu \pi \partial^\mu \pi) \cO + \tfrac{1}{4} \Mp^2 |\dot H| \tilde \mu^{- \Delta} ( - 2 \dot \pi + \partial_\mu \pi \partial^\mu \pi)^2 \cO  \ .
 \eeq
Here our operator $\cO$ couples to $\pi$ both linearly and quadratically. A bispectrum can therefore be induced by the power spectrum of $\cO$'s. Notice that this possibility arises already in the case in which only the first operator in (\ref{equ:doublemix0a}) is present.  However, as we explain in Appendix ~\ref{app:tuning}, this combination cannot give rise to a large $f_{NL}$. We therefore concentrate in the combination of the two operators. For simplicity we work just with the operators $\dot\pi$ and $\dot\pi^2$, and neglect the $(\partial_i\pi)^2$.  
 
 The induced bispectrum takes the following form
\bea\label{eq:bispectrumOO}
&&B(k_1,k_2,k_3) \equiv \langle\zeta_{\k_1}\zeta_{\k_2}\zeta_{\k_3}\rangle'= (\Mp^2 |\dot H|)^{3/2} \mu^{2- \Delta}\tilde\mu^{-\Delta}\ \times\\ \nonumber
&&\quad \langle \zeta_{\k_1}(\tau_0) \zeta_{\k_2}(\tau_0)  \zeta_{\k_3}(\tau_0)  \int  d \tau_1 d \tau_2\;a(\tau_0+i \tau_1)^4 a(\tau_0+i \tau_2)^4\ \times\  \\ \nonumber
&& \qquad   \dot\pi_{-\k_1}(\tau_0+i \tau_1) \dot\pi_{-\k_2}(\tau_0+i \tau_2)\dot\pi_{-\k_3}(\tau_0+i \tau_2) \rangle'\ \times \\ && \qquad  \langle   \cO_{\k_1} (\tau_0+i \tau_1)\cO_{-\k_1} (\tau_0+i \tau_2)\rangle'  + {\rm permutations} \nonumber \ .
\eea
A detailed understanding of the shape of this bispectrum will be the focus of remainder of this subsection.  
For our analysis it will be useful to work with the equivalent expression
\bea\label{equ:bispectrumOO}
&&B =\Big( \frac{\mu}{H}\Big)^{2 -  \Delta} \left(\frac{\tilde \mu}{H}\right)^{-\Delta}  \frac{\Delta^3_\zeta}{8} \frac{1}{k_1 k_2 k_3}\ \int d\tau_1\; (-i\tau_1 -\tau_0)^{\Delta-2} (1 + i k_1\tau_0 \tfrac{\tau_1}{|\tau_1|}  )e^{-k_1|\tau_1|} \\ \nonumber
&&\quad
 \int d\tau_2\; (-i\tau_2 -\tau_0)^{\Delta}\; (1 + i k_2\tau_0 \tfrac{\tau_2}{|\tau_2|}  )
 e^{-k_2|\tau_2|}\; (1 + i k_3\tau_0 \tfrac{\tau_2}{|\tau_2|}  )
 e^{-k_3|\tau_2|} \\
&&\quad\times \frac{(2\pi)}{4^{\Delta-1}}\frac{\Gamma(2-\Delta)}{\Gamma(\Delta)}\ \int d\omega\ e^{i\, \omega\,\tau_{12}}\; (\omega^2+k_1^2)^{\Delta-2}+{\rm permutations}\ , \nonumber
\eea
Here we have used (\ref{eq:real-2-point}) to express the $\langle \cO\cO\rangle'$ correlation function in Fourier space (and have
also introduced more compact notation $B \equiv B(k_1,k_2,k_3)$).  

\subsubsection{The Squeezed Limit}\label{sec:squeezedbispectrumOO}

Let us start by analyzing the squeezed limit. 
Understanding the scaling behavior of the bispectrum in the squeezed limit is instructive, both because it is an important signature of these models and because it is possible to perform analytically.  

Before beginning to analyse the formula (\ref{equ:bispectrumOO}), let us give a bit of intuition on how a contribution in the squeezed that is larger than in single clock inflation is generated. Let us concentrate in the squeezed limit $k_1\ll k_2\simeq k_3$. The reason why in single clock inflation there is a vanishingly small squeezed limit is that a long mode is locally unobservable, and so it cannot physically affect the correlation of two short modes. This determines the squeezed limit behavior. If we want to have a different squeezed limit, we need therefore to have locally observable long wavelength fluctuations. These are given by the long wavelength correlation of the $\cO$'s~\footnote{In the intuitive language described in~\cite{Senatore:2009cf} and in~\cite{LopezNacir:2011kk}, one can interpret the induced three-point function in the following way. The long-wavelength vacuum fluctuations of $\cO$'s generate a long $\pi$ mode by the coupling $\dot\pi\,\cO$, and affects the evolution of a short $\pi$ mode by its coupling $\dot\pi^2\cO$.  In the example of section \ref{sec:booo} it will instead be the correlated vacuum fluctuations of three $\cO$'s to generate directly a correlation between three $\pi$s though the mixing. The scaling in the squeezed limit can then be intuitively understood from the time-dependence of $\cO$ \cite{Dans}.}. This means that the leading contribution in the squeezed limit arises when the correlation function of the $\cO$'s is evaluated at momentum~$k_1$ (i.e. we can drop the permutations in equation \ref{eq:bispectrumOO}). Let us therefore concentrate of these terms. The $\omega$ integral is clearly peaked at $\omega\sim1/{\tau_{12}}$. Furthermore, because the integral has support only for $|\tau_{1,2}|\lesssim (k_{1,2})^{-1}$ respectively, $\tau_{12}^2\gtrsim 1/k_1^2$. We therefore can schematically write
\bea
&&B \sim\  \frac{1}{k_1 k_2 k_3}\ \int d\tau_1\; (-i\tau_1 -\tau_0)^{\Delta-2} (1 + i k_1\tau_0 \tfrac{\tau_1}{|\tau_1|}  )e^{-k_1|\tau_1|} \\ \nonumber
&&
 \quad\int d\tau_2\; (-i\tau_2 -\tau_0)^{\Delta}\; (1 + i k_2\tau_0 \tfrac{\tau_2}{|\tau_2|}  )
 e^{-k_2|\tau_2|}\; (1 + i k_3\tau_0 \tfrac{\tau_2}{|\tau_2|}  )
 e^{-k_3|\tau_2|} \ \  \tau_{12}^{-1}\ \tau_{12}^{2(2-\Delta)}\ . \nonumber
\eea

Let us consider the case $\Delta<2$ first.  First consider the $\tau_2$ integral. The integrand grows at least as fast as $|\tau_2|^{\Delta}$, which means that it is dominated by the largest possible values of $\tau_2$: $|\tau_2|\sim 1/k_2$. At this point, the integral in $\tau_1$ goes as $|\tau_1|^{\Delta-1}$ for $|\tau_1|\lesssim 1/k_2$, and as $|\tau_1|^{2-\Delta}$ for $|\tau_1|\gtrsim| \tau_2|$. The $\tau_1$ is dominated by the largest possible value of $\tau_1$ as well: $\tau_1\sim 1/k_1$. Putting these scaling together, we obtain the squeezed limit of the three point function to take the form
\bea\label{BscalingCFTOO}
B(k_1, k_2, k_3) &\propto& \frac{1}{k_1^3 k_2^3} \Big(\frac{k_1}{k_2} \Big)^{\Delta} \qquad {\rm for} \ \Delta \leq 2\ .
\eea
As advertised, this scaling depends directly on the conformal dimension of $\cO$.    

For the case $\Delta>2$, the $\tau_2$ integral behaves in the same way, being peaked at $|\tau_2|\sim 1/k_2$. Instead, the $\tau_1$ integral becomes a decreasing function of $|\tau_1|$ for $|\tau_1|\gtrsim |\tau_2|$. This means that even the $\tau_1$ integral is now peaked at $|\tau_1|\sim 1/k_2$. This gives the following squeezed limit
\bea\label{BscalingCFTOO2}
B(k_1, k_2, k_3) &\propto& \frac{1}{k_1^3 k_2^3} \Big(\frac{k_1}{k_2} \Big)^{2} \qquad {\rm for} \ \Delta > 2\ .
\eea
Notice that for $\Delta>2$ the integral has a UV divergence for $\tau_{12}\to 0$. We discuss this more in detail in Appendix ~\ref{app:tuning}, but here we just notice that the counterterm is $\dot\pi^3$, which has the same squeezed limit as (\ref{BscalingCFTOO2})~\cite{Senatore:2009gt}.

The distinctive squeezed limit that we find here and also in the next section has interesting consequences from the observational point of view. As noticed for the first time in~\cite{Dalal:2007cu} in the case non-Gaussianities of the local kind, when the squeezed limit of the bispectrum goes as $k_1^{-3} k_2^{-3}$, the bias of dark matter halos receives a contribution that scales as $1/k^2$ relative to the standard bias (see~\cite{Baldauf:2011bh} for a generalization to the full general relativistic setting). This is opened up the possibility to measure non-Gaussianities from the power spectrum of large scale structures. Analysis on the power spectra from current data in~\cite{Slosar:2008hx} have produced constraints comparable to the ones from CMB, while analysis of the bispectrum including the scale-dependent bias are expected to improve these limits even by about an order of magnitude~\cite{Baldauf:2010vn}. Models of quasi single field inflation have a more general squeezed limit, so that the scale dependence of the bias goes as $1/k^{\alpha}$ with $1/2\leq\alpha\leq2$. Non-Guassianities from our conformally coupled sector are able to generate a scale dependent bias in the different interval $0\leq\alpha=2-\Delta\leq 1$: in a sense, they allow us to fill the whole range. While a detection of the non-Gaussianity induced by these operators can be clearly within reach, the actual detection of these smaller values of $\alpha$ in the data seems to be quite hard; see~\cite{Norena:2012yi,Sefusatti:2012ye} for first forecasts using the scale dependent bias of the power spectrum of galaxies. It would be interesting to see if the prospects of detection will improve upon inclusion in the forecasts of the bispectrum of galaxies, as it was done for local non-Gaussianities in~\cite{Baldauf:2010vn}, as well as of  techniques that tend to reduce the cosmic variance~\cite{Seljak:2008xr}.

\subsubsection{The Shape and Amplitude of the Bispectrum}\label{sec:shapeOO}

In the previous subsection, we set up the calculation of the bispectrum and analyzed its squeezed limit analytically.  Here we will compute its shape numerically for various values of $\Delta$ and discuss the observational implications.

The bispectra that arise here are exactly scale invariant to the level of approximation we are considering.  As a result, the shape function defined in \cite{Babich:2004gb}\ determimes the signal-to-noise of the bispectrum.   It is given by $S(x_1,x_2) \equiv B(x_1, x_2, 1) x_1^2 x_2^2$, where $x_1=k_1/k_3$ and $x_2=k_2/k_3$;  the squeezed limit corresponds to $x_1 \to 0$ and $x_2 \to 1$.  In the examples we have discussed here, the shape functions scales as
\beq
\lim_{x_1 \to 0, x_2 \to 1} S(x_1, x_2) \propto x_1^{\Delta-1} \ .
\eeq
As a result, there is very little signal-to-noise in the squeezed limit for CFTs satisfying the unitarity bound $\Delta \geq 1$, apart from the introduction of a scale dependent bias.  For this reason, one should suspect that these models give equilateral or orthogonal type non-gaussianity.\footnote{In section \ref{timedep} we will exhibit unitary theories with more general scaling, including ones which simply generate local $f_{NL}$ to good approximation, with support in the squeezed limit.}
\begin{figure}[h!]
   \centering
       \includegraphics[width=0.45\textwidth]{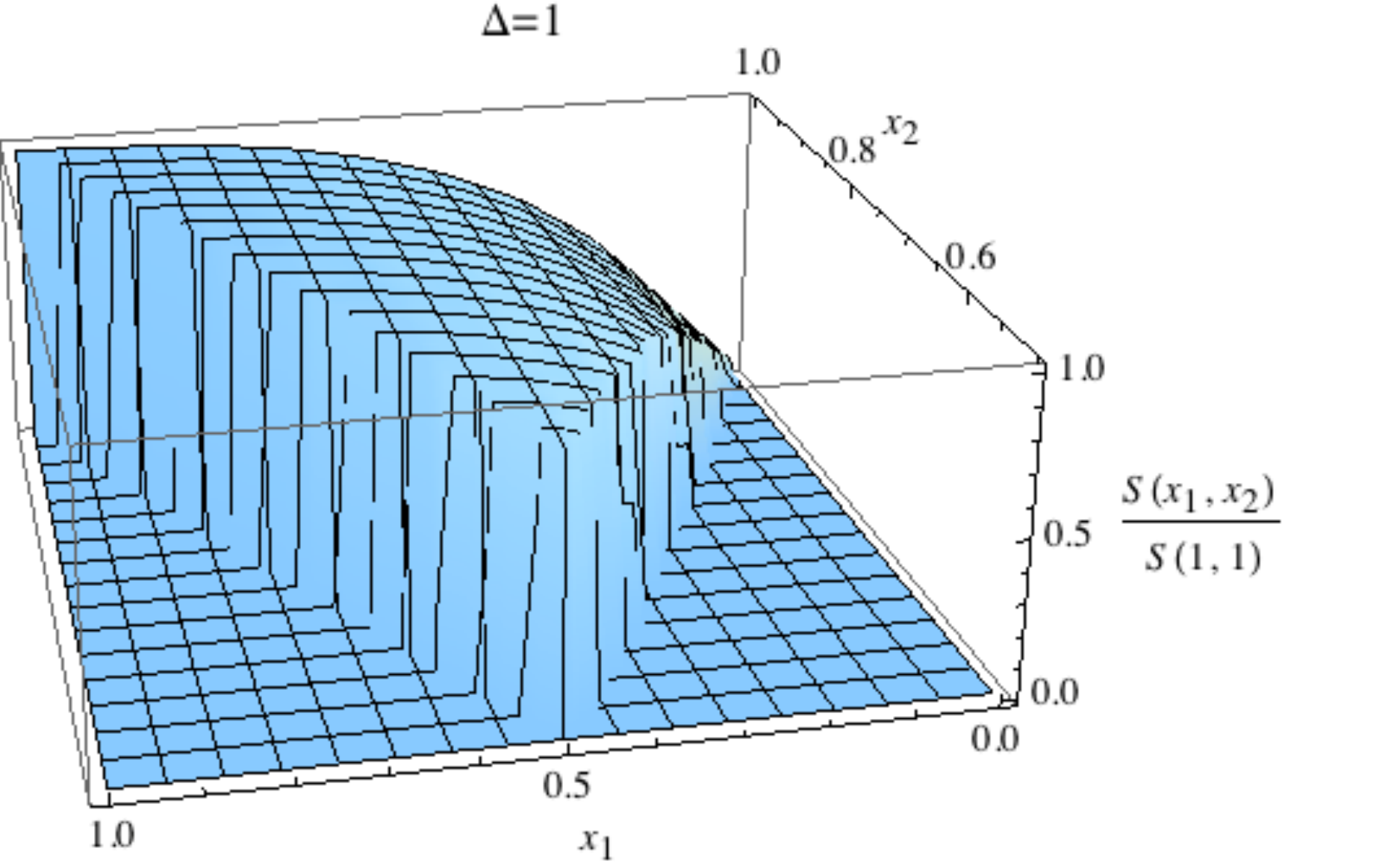}
              \includegraphics[width=0.45\textwidth]{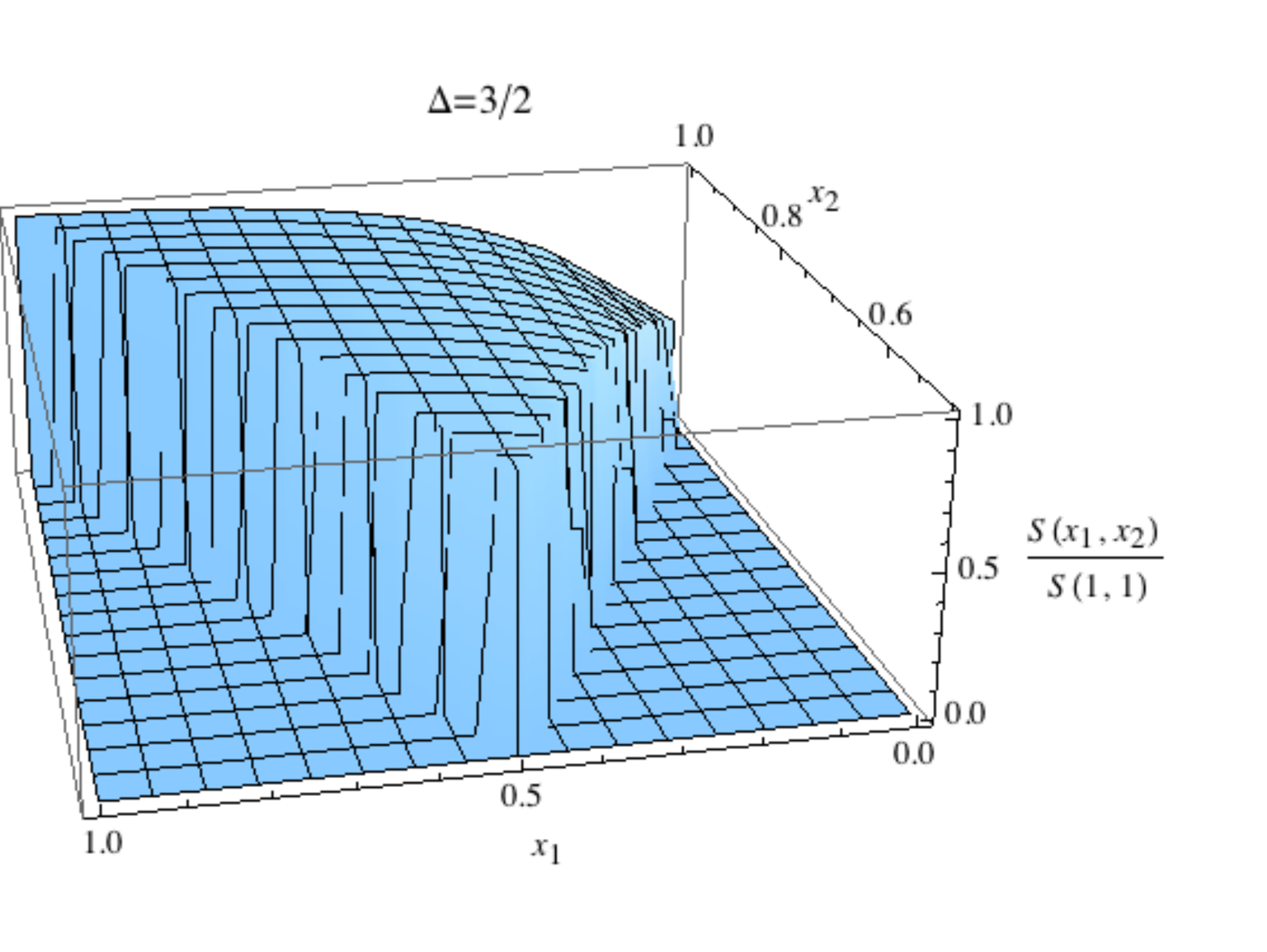}
   \caption{Numerically computed shape function, $S(x_1,x_2)$ evaluated for two values of $\Delta$}
  \label{fig:bispectrumdelta2OO}
\end{figure}
Evaluating the integrals in our expression  (\ref{equ:bispectrumOO}) for  bispectrum is not straightforward to do analytically in general, so we will compute it numerically. The two time integrals in (\ref{equ:bispectrumOO}) can be done analytically, and one is left to perform only the $\omega$ integral numerically~\footnote{We do not give here the result after the two time integrations, as it is just better to let your favorite Mathematica-like code do them for you.}.  A plot of the shape for $\Delta = 1$ 
and $3/2$ 
is shown in figure \ref{fig:bispectrumdelta2OO}.  It is difficult to discern by eye the distinction between the squeezed limit in this figure and the one of single clock inflation. This is an artefact of the plot related to the fact that the shapes starts to have the correct asymptotic squeezed limit only for $x_3\lesssim 10^{-1,-2}$. Such a delayed onset of the asymptotic squeezed limit makes it hard to read it in the plot, but we have verified it numerically.

Next, we will employ the optimal method for comparing shapes developed in \cite{Babich:2004gb}, to which we refer the reader for details.
The cosines, defined in equation (19) of \cite{Babich:2004gb}, between our shape for various values of $\Delta$ and the equilateral~\cite{Creminelli:2005hu}, orthogonal~\cite{Senatore:2009gt}, and local templates are shown in Table \ref{tab:cos1}.  We find that the results are consistent with equilateral or orthogonal shapes for the range of $\Delta$ allowed by the unitarity bound.

The appearance of the orthogonal shape for $\Delta \sim \tfrac{5}{4}$ is a special feature of the bispectrum generated by the $\langle\cO\cO\rangle$ two-point function, something which will not have a parallel in the next subsection when we consider the bispectrum generated by $\langle \cO \cO \cO\rangle$. 
While this paper was in preparation, we received the latest results from WMAP indicating a  2.45$\sigma$ hint of orthogonal non-Gaussianity \cite{Bennett:2012fp}.  Although this is not statistically significant with current data, it is worth noting the appearance of the orthogonal shape (with negligible support on the other templates) for a particular value $\Delta\sim \tfrac{5}{4}$ of the operator dimension.

\begin{table}[h!]
\caption{Cosine of shape with standard templates for operators of various dimensions. }
\label{tab:cos1}
\vspace{-0.5cm}
\begin{center}
\begin{tabular}{c c c c}
\toprule
\hspace{0.2cm} {\footnotesize \bf Dimension} & $\cos(S_\Delta, S_{\rm equilateral})$ & $\cos(S_\Delta, S_{\rm orthogonal})$   & $\cos(S_\Delta, S_{\rm local} )$  \\
\otoprule
$\Delta= 1$ &  0.95 & -0.07 & 0.45  \\
 \midrule
$\Delta= \tfrac{5}{4} $ &  0.05 & 0.78 & 0.59  \\
 \midrule
$\Delta= \tfrac{3}{2} $ &  0.93 & -0.15 & 0.45  \\
\bottomrule
\end{tabular}
\end{center}
\end{table}  

\begin{figure}[h!]
   \centering
       \includegraphics[scale =0.6]{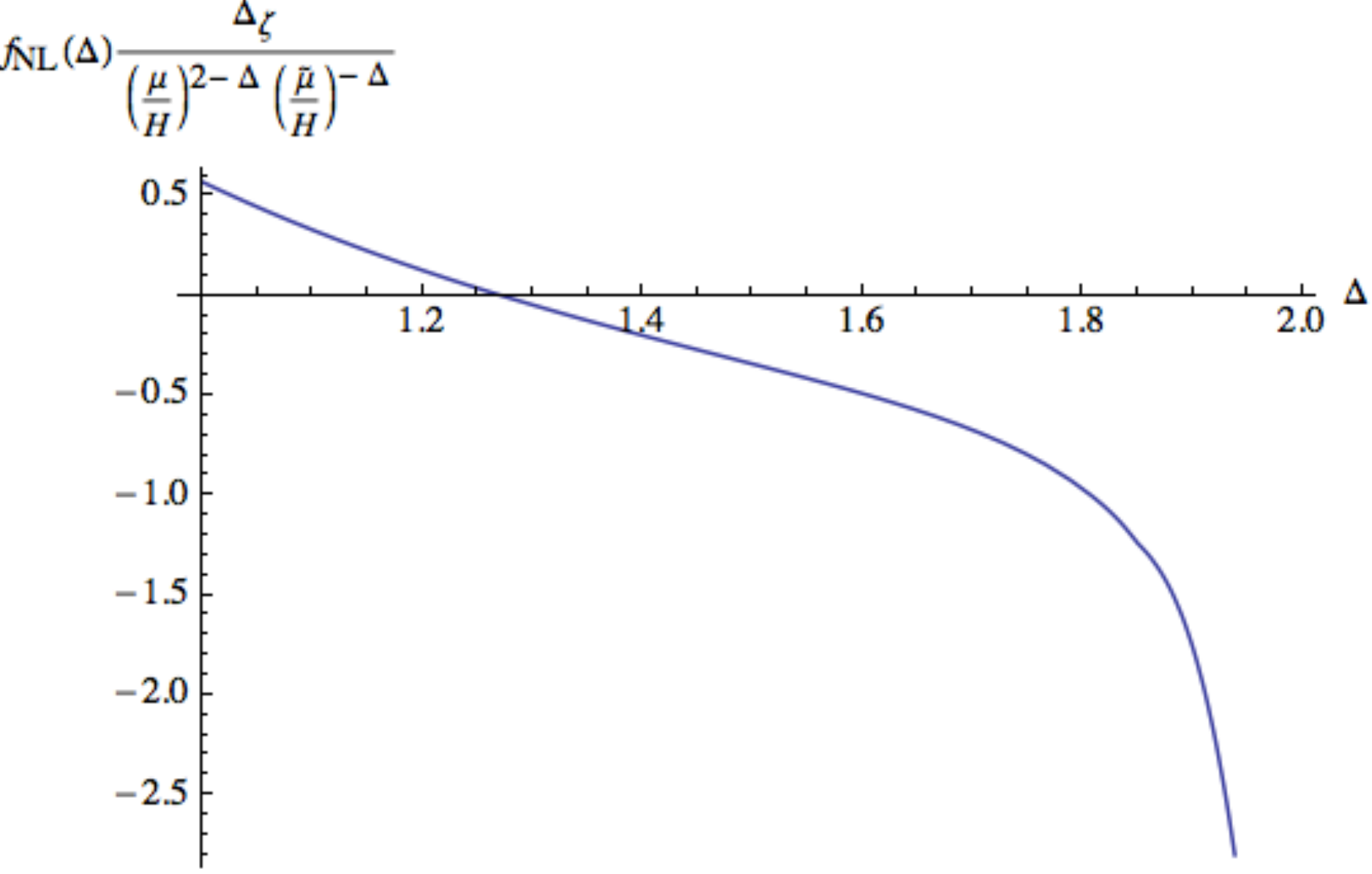}
   \caption{$f_{NL}$ as a function of $\Delta$.}
  \label{fig:fnlOO}
\end{figure}

The value of $f_{\rm NL}$ can be defined in the standard way:
\beq
f_{\rm NL} \Delta_\zeta=\frac{5}{18} \frac{\langle\zeta_k\zeta_k\zeta_k\rangle'}{\Delta_\zeta^{3}}
\eeq
From the interaction terms leading to our general expression (\ref{equ:bispectrumOO}), it is easy to read off  that $f_{NL}$ scales as $\left(\mu/H\right)^{2-\Delta}\left(\tilde\mu/H\right)^{-\Delta}$.  
We plot the value of $f_{\rm NL} \Delta_\zeta/\left(\left(\mu/H\right)^{2-\Delta}\left(\tilde\mu/H\right)^{-\Delta}\right)$ as a function of $\Delta$ in Fig.~\ref{fig:fnlOO}.

We see the expected divergence as $\Delta\to 2$, discussed above in the in the previous subsection and appendix A.
Note that the vanishing value of $f_{\rm NL}$ around $\Delta \simeq 5/4$ does not indicate that the non-gaussianity vanishes, just that it goes to zero in the equilateral limit.  We see this explicitly from full shape for $\Delta\simeq5/4$, which is similar to the orthogonal template.

  \begin{figure}[h!]
   \centering
       \includegraphics[scale =.55]{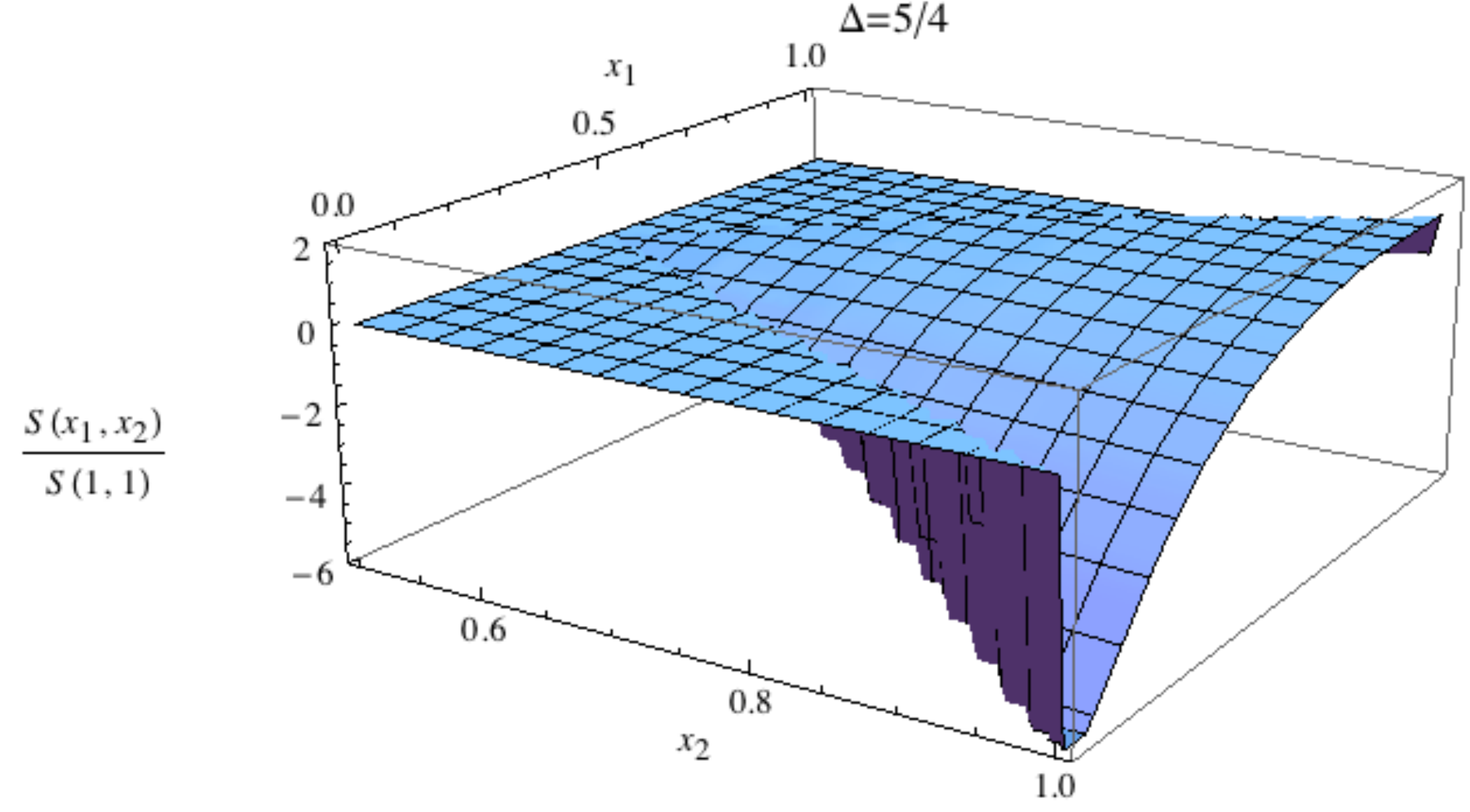}
   \caption{the shape function for $\Delta=5/4$.}
  \label{fig:ShapeDeltaOO}
\end{figure} 

Finally, let us briefly comment on the size of $f_{\rm NL}$ in a simple example like (\ref{dimsix}).  If both $\mu$ and $\tilde \mu$ were generated at a common scale, $M_*$, it is natural to expect $\mu^{2 -\Delta} = \dot \phi / M_*^{\Delta}$ and $\tilde \mu^{-\Delta} = \dot \phi^2 / M_*^{4+\Delta}$.  However, for $\Delta < 2$, the constraint $\mu < H$ implies that $\left(\mu/H\right)^{2-\Delta}\left(\tilde\mu/H\right)^{-\Delta}  \lesssim \Delta_\zeta^{\tfrac{4}{\Delta} -1}$ and therefore the bispectrum generated by $\langle\cO\cO\rangle$ satisfies $f_{\rm NL} \lesssim 1$.    Achieving large $f_{\rm NL}$ is still possible but requires multiple scales (e.g. $M_*$ and $\tilde M_*$) or a different UV completions altogether.

\subsection{Bispectrum from $\langle\cO\cO\cO\rangle$}\label{sec:booo}

In this subsection, we will consider the bispectrum of $\pi$ that is generated by ${\cal H}_{\rm int} = \tfrac{1}{\sqrt{2}} \mu^{2-\Delta} \dot \pi_c \cO$ and the three point function of $\cO$.  The leading contribution to the bispectrum is given by
\bea\label{eq:bispectrum}
B &=&- \Mp^3 |\dot H|^{3/2} \mu^{6-3 \Delta}\ \times\\ \nonumber
&&\quad \langle \zeta_{\k_1}(\tau_0) \zeta_{\k_2}(\tau_0)  \zeta_{\k_3}(\tau_0)  \int  d \tau_1 d \tau_2 d \tau_3 \;a(\tau_0+i \tau_1)^4 a(\tau_0+i \tau_2)^4a(\tau_0+i \tau_3)^4\ \times\  \\ \nonumber
&& \quad   \dot\pi_{-\k_1}(\tau_0+i \tau_1) \dot\pi_{-\k_2}(\tau_0+i \tau_2)\dot\pi_{-\k_3}(\tau_0+i \tau_3) \rangle'\ \langle \cO_{\k_1} (\tau_0+i \tau_1)\cO_{\k_2} (\tau_0+i \tau_2)\cO_{\k_3} (\tau_0+i \tau_3) \rangle' \ ,
\eea
where, again, $B \equiv B(k_1, k_2, k_3) \equiv \langle\zeta_{\k_1}\zeta_{\k_2}\zeta_{\k_3}\rangle'$.  As in the previous subsection, we will start by considering the squeezed limit  before discussing the full shape.

\subsubsection{The Squeezed Limit}\label{sec:squeezedbispectrum}

The calculation of the bispectrum simplifies in the squeezed limit, $k_1 \ll k_2, k_3$, if this limit also corresponds in the OPE limit of the 3 point function of $\cO$ in position space: $|\x_2 - \x_3| \ll |\x_1 - \x_2| \sim |\x_1 -\x_3|$ and similarly $|\tau_2 - \tau_3| \ll |\tau_1 - \tau_2| \sim |\tau_1 -\tau_3|$.  To estimate when these limits coincide, let us write
\bea\label{equ:bispectrum}
B =-\Big( \frac{\mu}{H}\Big)^{6 - 3 \Delta} \frac{\Delta^3_\zeta}{8 } \frac{1}{k_1 k_2 k_3}&& \Big( \prod_{i = 1}^{3} \int d\tau_i (-i\tau_i -\tau_0)^{\Delta-2} (1 + i k_i\tau_0 \tfrac{\tau_i}{|\tau_i|}  )
 e^{-k_i|\tau_i|}\Big) \\
&&\times \int  d^3 x_{13} d^3 x_{23} \frac{C e^{i \k_1 \cdot \x_{13} + i \k_2 \cdot \x_{23} }}{(|\x_{13 }- \x_{23} |^2 + \tau^2_{12})^{\tfrac{\Delta}{2}} (x_{23}^2 + \tau^2_{23})^{\tfrac{\Delta}{2}} (x_{13}^2 + \tau^2_{13})^{\tfrac{\Delta}{2}} } \ .\nonumber
\eea
The integrals over $x_{13}$ and $x_{23}$ will receive most of their support from $ x_{23} \lesssim k_2^{-1}$ and $x_{13} \lesssim k_1^{-1}$ due to the oscillations at larger values.  Similarly, the integrals over $\tau_i$ are exponentially suppressed unless $|\tau_i| \lesssim k_i^{-1}$.  In the squeezed limit, $k_1 \ll k_2, k_3$.  As a result, the integral receives support from $\tau_1 \gg \tau_2,\tau_3$ and $x_1 \gg x_2 , x_3$.  If these configurations dominate the integral, then we can use the OPE limit of the three point function to approximate the squeezed limit.  

We want to determine if the integral receives its dominant contribution in the OPE limit.  To do so, we may assume that $x_2, x_3, \tau_2,\tau_3 \sim k_{2}^{-1} \sim k_3^{-1} \ll k_1^{-1}$ as larger values are suppressed.  Now we will integrate over $x_{13}$ at fixed $\tau_1$, assuming $x_{12} \sim x_{13}$.  This assumption is reliable is the integral is dominated by $x_1 \gg x_2, x_3$.  For $x_{13} \ll \tau_{13}$ the integral scales as $x_{13}^3$ and therefore it is dominated by the largest values of $x_{13}$.  For $x_{13} \gg \tau_{13}$ the integral scales as $x_{13}^{3- 2\Delta}$.  For $\Delta< \tfrac{3}{2}$, the largest values of $x_{13}$ still dominate up to the cutoff, yielding $x_{13 } \sim k_1^{-1}$.  Having performed the $x_{13}$ integral, we now perform the integral over $\tau_1$ to find $\tau_{1} \sim k_1^{-1}$.  As a result, when $\Delta < \tfrac{3}{2}$ the integrals are dominated where the OPE limit of the CFT is applicable.

For $\Delta > \tfrac{3}{2}$, the integral over $x_{13}$ is peaked at $x_{13} \sim \tau_{13}$.  Now we perform the integral $\tau_1$ with $x_{13} \sim \tau_{13}$, which scales as $\int d\tau_1 |\tau_1|^{1-\Delta} \sim \tau_{1}^{2-\Delta}$.  For $\Delta < 2$, this integral is peaked at $\tau_1 \sim k_1^{-1}$ where the integral is cutoff by the exponential.  In this case, the OPE limit is again applicable.  However, for $\Delta > 2$, the integral is peaked around $\tau_1 \sim k_2^{-1}$ where the OPE limit does not apply~\footnote{In a correlation function in a CFT involving more than two operators inserted at various points, the OPE between two operators inserted at two points is convergent as long as there is ball that contains those two points but no other one.}.  Schematically, these results imply
\bea\label{BscalingCFT}
B &\propto& \frac{1}{k_1^3 k_2^3} \Big(\frac{k_1}{k_2} \Big)^{\Delta} \qquad {\rm for} \ \Delta \leq 2\\
&\propto& \frac{1}{k_1 k_2^5} =  \frac{1}{k_1^3 k_2^3} \Big(\frac{k_1}{k_2} \Big)^{2} \qquad \qquad \quad{\rm for} \ \Delta \ge 2 \ .
\eea

\begin{figure}[h!]
   \centering
       \includegraphics[scale =0.55]{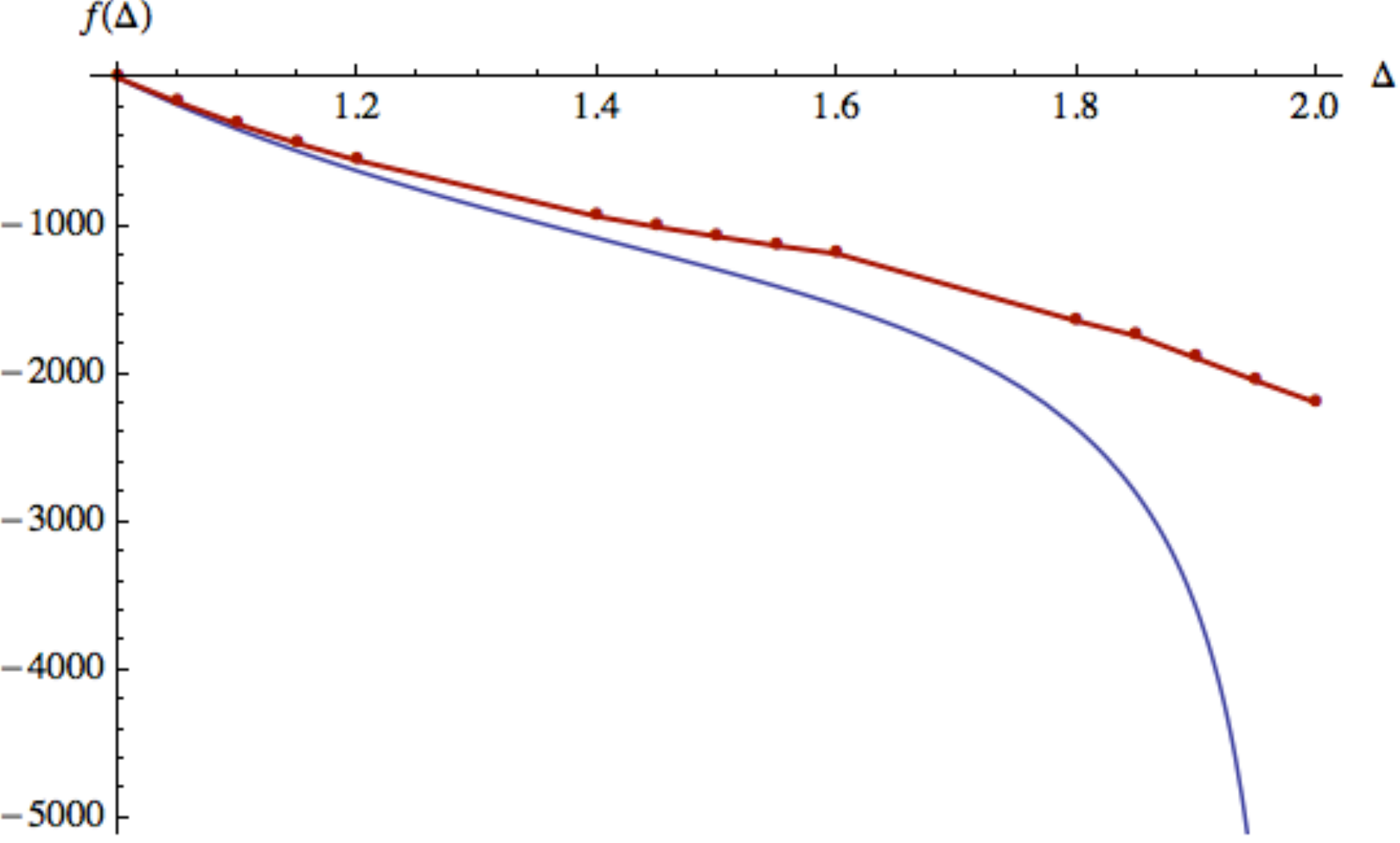}
   \caption{Numerically computed $f(\Delta)$.}  
  \label{fig:fsqueezed}
\end{figure}

Now let us compute the $\Delta < 2$ case more carefully for the squeezed limit.  We have established that the OPE limit of the three-point function is where the integral is dominated, so we can take
\bea
B =-\Big( \frac{\mu}{H}\Big)^{6 - 3 \Delta} \frac{\Delta^3_\zeta}{8}&& \frac{1}{k_1 k_2 k_3}\Big( \prod_{i = 1}^{3} \int d\tau_i (-i\tau_i -\tau_0)^{\Delta-2} (1 + i k_i \tau_0 \tfrac{\tau_i}{|\tau_i|}  )
 e^{-k_i|\tau_i|}\Big) \\&&\times \int  d^3 x_{13} d^3 x_{23} \frac{C e^{i \k_1 \cdot \x_{13} + i \k_2 \cdot \x_{23} }}{(\x_{13 }^2 + \tau^2_{13})^{\Delta} (x_{23}^2 + \tau^2_{23})^{\tfrac{\Delta}{2}} } \nonumber \\
=-\Big( \frac{\mu}{H}\Big)^{6 - 3 \Delta} \frac{\Delta^3_\zeta}{8} \frac{1}{k_1 k_2 k_3}&&\Big( \prod_{i = 1}^{3} \int d\tau_i (-i\tau_i -\tau_0)^{\Delta-2} (1 + i k_i \tau_0 \tfrac{\tau_i}{|\tau_i|}  )
 e^{-k_i|\tau_i|}\Big)  \\
\times (2\pi)^4 \frac{\Gamma(2-\Delta)}{2^{2\Delta-2} \Gamma(\Delta)}&&\frac{\Gamma(2-\tfrac{\Delta}{2})}{2^{\Delta-2} \Gamma(\tfrac{\Delta}{2})} \int \frac{d \omega_1 d\omega_2}{(2\pi)^2}  e^{i (\omega_1\tau_{13} + \omega_2 \tau_{23} )}(\omega_1^2+ k_1^2)^{\Delta-2} ( \omega_2^2+k_2^2)^{\tfrac{\Delta}{2}-2} \nonumber 
\eea
We can evaluate the $\tau_i$ integrals using 
\bea\label{equ:j}
{\cal J}(k, \omega, \tau_0)&\equiv&\int_{-\infty}^{\infty} d\tau (-i \tau - \tau_0)^{\Delta-2} (1 + i k\tau_0 \tfrac{\tau_i}{|\tau_i|}  ) e^{- k |\tau| +i \omega \tau}  \\ 
&=&\frac{ i e^{-(i k \tau_0 +\tau_0 \omega + \tfrac{i \pi}{2} \Delta)}}{ (k^2 +\omega^2)^{\Delta} }\Big[ e^{2 i k\tau_0 + i \pi \Delta} (k+i \omega) (k-i \omega)^{\Delta} (i + k \tau_0) \Gamma[\Delta-1,i \tau_0 (k+i\omega)] \nonumber \\&& -  (k-i \omega) (k+i \omega)^{\Delta} (-i + k \tau_0) \Gamma[\Delta-1,-i \tau_0 (k-i\omega)] \Big]  \ .
\eea
The squeezed bispectrum is then given by
\bea
B = - \Big( \frac{\mu}{H}\Big)^{6 - 3 \Delta} \frac{\Delta^3_\zeta}{8} \frac{C \kappa}{k_1 k_2 k_3}  \int  \frac{d \omega_1 d\omega_2}{(2\pi)^2} && \hskip -8pt  (\omega_1^2+ k_1^2)^{\Delta-2} ( \omega_2^2+k_3^2)^{\tfrac{\Delta}{2}-2} \nonumber \\ \times && \hskip  -24pt\  \ { \cal J}(k_1, \omega_1, \tau_0) {\cal J}(k_2, \omega_2, \tau_0){\cal J}(k_3, -\omega_1 - \omega_2, \tau_0) \ ,
\eea
where
\beq
\kappa = (2\pi)^4 \frac{\Gamma(2-\Delta)}{2^{2\Delta-2} \Gamma(\Delta)}\frac{\Gamma(2-\tfrac{\Delta}{2})}{2^{\Delta-2} \Gamma(\tfrac{\Delta}{2})} 
\eeq
In the squeezed limit, the integral is dominated by $\omega_2 \sim k_2$ and $\omega_1 \sim k_1$ so we may set $\omega_1+\omega_2 \to \omega_2$ and $k_3 \sim k_2$.  The remaining $k_{1,2}$ is dependence determined in the $\tau_0 \to 0$ by rescaling $\omega_i$ by $k_i$, such that the integrals are $k$-independent.  The squeezed limit is then determined by
\bea
B =&& \Big( \frac{\mu}{H}\Big)^{6 - 3 \Delta} \Delta^3_\zeta \frac{ \alpha  }{k_1^{3-\Delta} k_2^{3+\Delta}} f(\Delta)
\eea
where the function $f(\Delta)$ is shown in Fig. \ref{fig:fsqueezed} in the blue un-dotted line. In the last passage, we have parametrized $C=5 \alpha (\Delta-1)$. This parametrization originates from the fact that consistency of the CFT under crossing symmetry~\footnote{This is sometimes referred to as `bootstrap' or as `OPE associativity constraint'.}, requires the following numerically-found upper bound $C\leq 5\, \alpha\,  (\Delta-1)$, where $\alpha$ is bounded above by a number numerically close to one~\cite{Caracciolo:2009bx,Poland:2011ey}. This result is relatively intuitive, as $\Delta=1$ represents a free theory, and the bound just quoted implies that the $C$ is continuous in the limit $\Delta\to 1$.     

The rapid growth of $f(\Delta)$ near $\Delta=2$ is not physical, but represents the increase in the error we are making by using only the leading term in the OPE.  The breakdown in the OPE at $\Delta = 2$ appears as a logarithmic divergence in $f(\Delta)$ at $\Delta =2$. This divergence is removed if we compute the squeezed limit using the generally valid formula that we explain in the next section. We plot this function in Fig.~\ref{fig:fsqueezed} in the red dotted line, where we see that the limit $\Delta\to 2$ is smooth and the contribution remains finite. 
We expect that the difference between the analytic and the numerical functions for $\Delta\to 1$ is
a numerical uncertainty.

\subsubsection{The Shape and Amplitude of the Bispectrum}\label{sec:shape}

Now let us consider the shape of the bispectrum as a function of $\Delta$. As we did in Sec.~\ref{sec:shapeOO}, we will numerically plot the shape function, $S(x_1, x_2)$, and compare it to the standard templates using the cosines defined in equation (19) of  \cite{Babich:2004gb}.  We will also determine the value of $f_{NL}$, finding it to be naturally substantial while at the same time bounded in an interesting way by limits on the size $C$ of CFT three-point functions found using crossing symmetry.   

\begin{figure}[h!]
   \centering
       \includegraphics[width=0.45\textwidth]{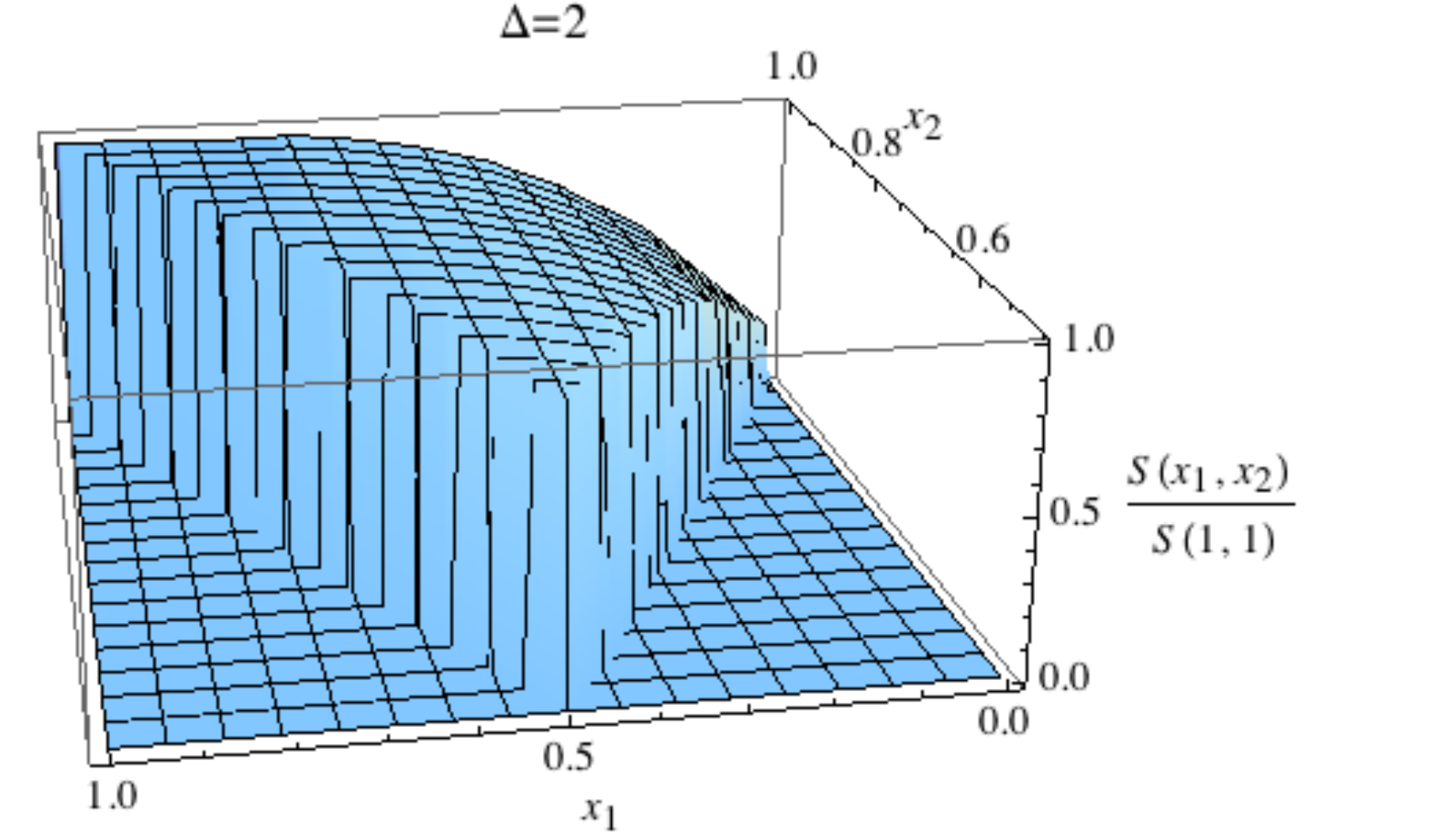}
              \includegraphics[width=0.45\textwidth]{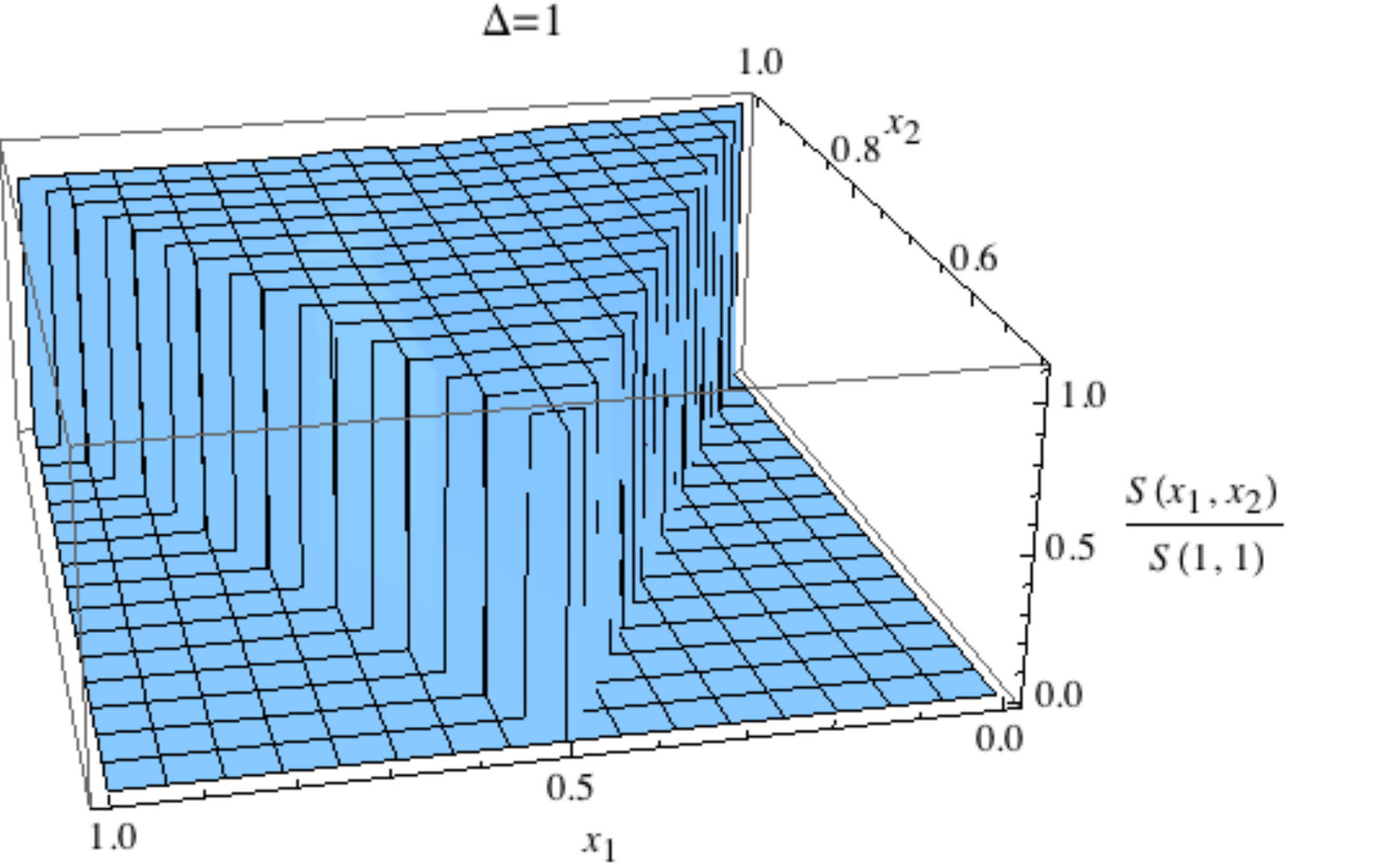}
   \caption{Numerically computed shape function, $S(x_1,x_2)$ evaluated for $\Delta = 2$ (top) and $\Delta = 1$ (bottom)}
  \label{fig:bispectrumdelta2}
\end{figure}
Computing the bispectrum analytically is not straightforward, but it can be computed numerically.  To speed up the integration, one can perform some of the integrations analytically, as we discussion in Appendix \ref{app:details}.  Aside from the scaling in the squeezed limit, the shapes do not change dramatically as a function of $\Delta$.  A plot of the shape for $\Delta = 1$ and $2$ is shown in Fig. \ref{fig:bispectrumdelta2}.  

The cosine, defined in \cite{Babich:2004gb}, between our shape for $\Delta = 2$ and the equilateral~\cite{Creminelli:2005hu}, orthogonal~\cite{Senatore:2009gt}, and local templates are shown in Table \ref{tab:cos2}.  The cosine varies little as a function of $\Delta$ for all values consistent with the unitarity bound, $\Delta \geq 1$.  As a result, we find that our bispectrum is largely equilateral in shape. 

\begin{table}[h!]
\caption{Cosine of shape with standard templates for operators of various dimensions. }
\label{tab:cos2}
\vspace{-0.5cm}
\begin{center}
\begin{tabular}{c c c c}
\toprule
\hspace{0.2cm} {\footnotesize \bf Dimension} & $\cos(S_\Delta, S_{\rm equilateral})$ & $\cos(S_\Delta, S_{\rm orthogonal})$   & $\cos(S_\Delta, S_{\rm local} )$  \\
\otoprule
$\Delta= 1$ &  0.94 & -0.11 & 0.47  \\
 \midrule
$\Delta= 2 $ &  0.90 & -0.25 & 0.56  \\
\bottomrule
\end{tabular}
\end{center}
\end{table}

\begin{figure}[h!]
   \centering
       \includegraphics[scale =1]{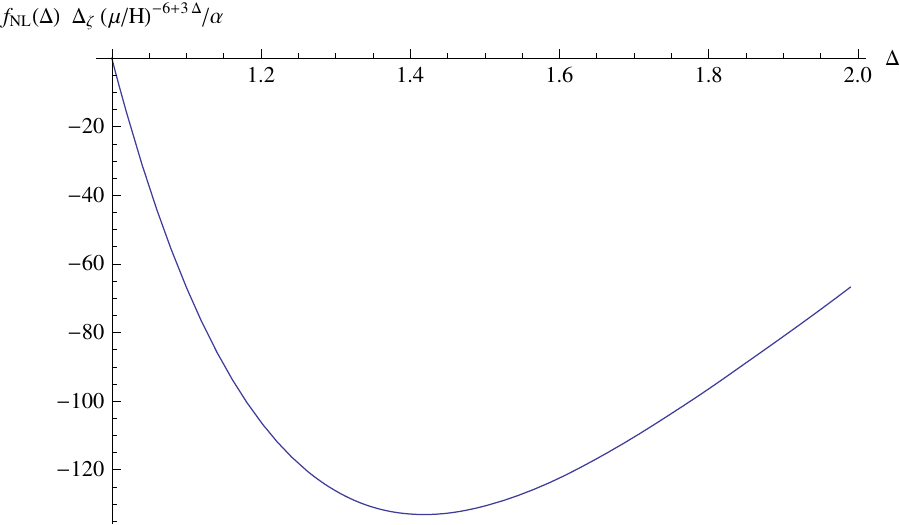}
   \caption{$f_{NL}$ as a function of $\Delta$.}
  \label{fig:fnl}
\end{figure}

The amplitude of the bispectrum,  $f_{NL}$, clearly scales as $C \left(\mu/H\right)^{6-3\Delta}$ since it is proportional to the CFT three-point function and three insertions of the mixing interaction. As discussed above, $C$ is bounded as $C\leq 5\, \alpha\,  (\Delta-1)$, where the coefficient $\alpha$ is bounded above by a number numerically close to one~\cite{Caracciolo:2009bx,Poland:2011ey}.
We plot the value of $f_{NL} \Delta_\zeta/(\mu/H)^{6-3\Delta}/\alpha$ as a function of $\Delta$ in Fig.~\ref{fig:fnl}.  Since $\alpha\lesssim 1$, the curve can be thought as an upper bound to the value of $f_{NL}$. We see that for $\Delta\to 1$, the $f_{NL}$ induced by $\langle\cO\cO\cO\rangle$ goes to zero.
However, $C$ can be of order one already as soon as $\Delta\gtrsim 1.2$, so large values of $f_{NL}$ easily fit within the parameter range of consistent theories.   

In particular, for $C$ of order 1 the amplitude of the (nearly equilateral) non-Gaussianity 
gives us a sensitive probe of higher dimension operators such as (\ref{dimsix}) suppressed by a high scale $M_*$ which can be much larger than $H$.  Similar remarks apply for all the cases we analyze in this paper.  

\subsection{The Collapsed Limit of the Tri-Spectrum}

Another important signal of these models is the collapsed limit of the tri-spectrum.  The behavior in this limit is often parameterized for local-type non-gaussanity in terms of $\tau_{\rm NL}$ which is defined as 
\beq
\tau_{\rm NL} = \frac{1}{4} \lim_{|\k_{1} +\k_{2}| \to 0}\frac{ \langle \zeta_{\k_1} \zeta_{\k_2} \zeta_{\k_3} \zeta_{\k_4} \rangle'}{{\cal P}_{\zeta}(k_1) {\cal P}_{\zeta}(k_3) {\cal P}_{\zeta} (|\k_{1}+ \k_{2}| ) } \ .
\eeq
If $\tau_{\rm NL} > (\tfrac{6}{5} f_{\rm NL} )^2$ (the $\tfrac{6}{5}$ is a result of the conventions of $f_{\rm NL}$ and $\tau_{\rm NL}$), then more than one field must contribute to $\zeta$ \cite{SY,Sugiyama:2011jt, Lewis:2011au, Smith:2011if, Quasi4,Kehagias:2012pd}.  Such models also have important consequences in large scale structure, where they lead to scale-dependent {\it stochastic} bias \cite{Smith:2010gx, Baumann:2012bc}.  These result generalize straightforwardly to tri-spectra that scale as any inverse power of $|\k_{1}+ \k_{2}|$ in the limit $|\k_{1} +\k_{2}| \to 0$ (see e.g.~\cite{Quasi4}).  Our models include such generalizations so we will define $\tau_{\rm NL}$ to be
\beq\label{equ:taudef}
\tau_{\rm NL} = \frac{1}{4} \lim_{|\k_{1} +\k_{2}| \to 0}  \frac{(k_1 k_3)^{\Delta_i}}{|\k_{1}+ \k_{2}|^{2 \Delta_i} } \frac{ \langle \zeta_{\k_1} \zeta_{\k_2} \zeta_{\k_3} \zeta_{\k_4} \rangle'}{{\cal P}_{\zeta}(k_1) {\cal P}_{\zeta}(k_3) {\cal P}_{\zeta} (|\k_{1}+ \k_{2}| ) }  \ ,
\eeq
where $\Delta_i$ is a parameter that is determined from the leading contribution to this limit.  In Section \ref{sec:4otri}, we will find that $\Delta_i$ is the lowest dimension appearing in the OPE of $\cO(x) \cO(0)$.  

Whenever extra degrees of freedom contribute significantly to the squeezed limit of the bispectrum without contributing significantly to the power spectrum, a large contribution to  $\tau_{NL}$ is expected \cite{Quasi4}.  Specifically, under such circumstances, one can expect to find $\tau_{\rm NL} \gg (\tfrac{6}{5} f_{\rm NL})^2$, where $f_{NL}$ here is the one induced by $\langle\cO\cO\cO\rangle$.  The basic intuition is that generating $f_{\rm NL}$ requires three $\dot\pi\cO$ mixing interactions, whereas $\tau_{\rm NL}$ gets a leading contribution from four mixing interactions combined with the CFT four-point function.  As a result, the tri-spectrum is enhanced by two inverse powers of the coupling controlling the mixing.  In this section, we will confirm this intuition for our CFT examples.

We will discuss two concrete possibilities for a large tri-spectrum.  First, we will consider the tri-spectrum induced by the nonlinear coupling of $\cO$ to $\dot \pi^2$ that we discussed in Section \ref{sec:boo}.  The full tri-spectrum is computable in this case, but we will focus on the collapsed limit.  Second, we will consider the tri-spectrum induced by the linear mixing interaction used in Section \ref{sec:booo}.  Here the tri-spectrum is determined by  $\langle \cO\cO \cO \cO\rangle$ and is not determined by conformal invariance alone.  The collapsed limit can be understood in terms of the OPE of $\cO$. 

\subsubsection{The $\langle \cO\cO\rangle$-induced Tri-Spectrum}

From the discussion in Appendix \ref{app:tuning}, it is possible that the only interaction term is ${\cal H} =  \dot\pi_c^2\cO/2\tilde\mu^\Delta$. Under these circumstances, the bispectrum would vanish, but a large tri-spectrum could be generated as follows.  We will define ${\cal Y} \equiv \langle \zeta_{\k_1} \zeta_{\k_2} \zeta_{\k_3} \zeta_{\k_4} \rangle'$, and we get
\bea
{\cal Y} &=& \tilde\mu^{- 2 \Delta} \Mp^4 |\dot H|^{2}
\int  \frac{ d\tau_1  }{H^3 (- i \tau_1 - \tau_0 )^3} 
\int  \frac{ d\tau_2  }{H^3 (- i \tau_2 - \tau_0 )^3} \\ \nonumber
&\times&\langle \zeta_{\k_1}(\tau_0) \pi'_{-\k_1}(\tau_1 )\rangle' 
\langle \zeta_{\k_2}(\tau_0) \pi'_{-\k_2}(\tau_1 )\rangle' 
\langle \zeta_{\k_3}(\tau_0) \pi'_{-\k_3}(\tau_2 )\rangle' 
\langle \zeta_{\k_4}(\tau_0) \pi'_{-\k_4}(\tau_2 )\rangle' \\ \nonumber
&\times& \langle\cO_{\k_1+\k_2}(\tau_1)\cO_{-\k_1-\k_2}(\tau_2)\rangle'  \nonumber
\eea
where again $\langle\dots\rangle'$ indicates dropping the momentum conserving delta function.  It is straightforward to plug in the two point functions of $\pi$ and $\cO$ to find
\bea\label{equ:trispectrumOO}
&&{\cal Y} = \left(\frac{\tilde \mu}{H}\right)^{-2 \Delta}  \frac{\Delta^4_\zeta}{16} \frac{1}{k_1 k_2 k_3 k_4}\ \int d\tau_1\, (-i\tau_1 -\tau_0)^{\Delta} (1 + i k_1\tau_0 \tfrac{\tau_1}{|\tau_1|}  )(1 + i k_2\tau_0 \tfrac{\tau_1}{|\tau_1|}  )e^{-(k_1+k_2)|\tau_1|} \nonumber \\ 
&&\quad
 \int d\tau_2\; (-i\tau_2 -\tau_0)^{\Delta}\; (1 + i k_3\tau_0 \tfrac{\tau_2}{|\tau_2|}  )
\; (1 + i k_4\tau_0 \tfrac{\tau_2}{|\tau_2|}  )
 e^{-(k_3+k_4)|\tau_2|} \\
&&\quad\times \frac{(2\pi)}{4^{\Delta-1}}\frac{\Gamma(2-\Delta)}{\Gamma(\Delta)}\ \int d\omega\ e^{i\, \omega\,\tau_{12}}\; (\omega^2+|\k_1+\k_2|^2)^{\Delta-2}+{\rm permutations}\ . \nonumber
\eea
From here, it is straightforward to compute the full tri-spectrum numerically but we will not show the result here.

In order to gain more insight into the form of the tri-spectrum, we will consider the collapsed limit, $|\k_1 + \k_2 | \to 0$, analytically.  Because the exponential suppression, $\tau_1 < k_1^{-1}$ and $\tau_2 < k_3^{-1}$.  Using $|\k_1 + \k_2 | \ll k_1, k_3$, we see that for $\Delta < \tfrac{3}{2}$ the $\omega$ integral is dominated by $\omega \sim |\k_1 +\k_2|$.  Therefore, we can ignore the factor of $e^{i \omega \tau_{12}}$ and compute all three integrals analytically.  As a result, we find
\bea
{\cal Y} &\to& \left(\frac{\tilde \mu}{H}\right)^{-2 \Delta}  \Delta^4_\zeta \frac{\pi^{3/2}}{4^\Delta}\Delta(1+\cos(\pi\Delta))\Gamma(3/2-\Delta)\Gamma(1+\Delta) \frac{|\k_1 +\k_2|^{2\Delta - 3}}{k_1^{3+\Delta} k_3^{3+\Delta}}  ,
\eea
We see that tri-spectrum scales as $ k_1^{-3-\Delta}k_3^{-3-\Delta}|\k_1 +\k_2|^{2\Delta -3}$ in the limit $|\k_1 + \k_2 | \to 0$. For $\Delta>3/2$, the tri-spectrum scales as $k_1^{6-\Delta}k_3^{6-\Delta} ({\rm Max}[k_3,k_1])^{2\Delta-3}$.  Using $\Delta_i = \Delta$ in (\ref{equ:taudef}), we see that $\tau_{\rm NL} \propto \left( \frac{\tilde \mu}{H}\right)^{-2 \Delta} \Delta_{\zeta}^{-2}$ and is potentially in the measurable range.  Furthermore, because the bispectrum vanishes at tree level and radiative corrections can be small, as discussed in Appendix \ref{app:tuning}, this can be the leading source of non-gaussanity.

\subsubsection{The $\langle \cO\cO \cO \cO\rangle$-induced Tri-Spectrum}\label{sec:4otri}
Now let us repeat the calculation of the tri-spectrum (again defining ${\cal Y} \equiv \langle \zeta_{\k_1} \zeta_{\k_2} \zeta_{\k_3} \zeta_{\k_4} \rangle'$) using only the interaction ${\cal H}_{\rm int} = \tfrac{1}{\sqrt{2}} \mu^{2-\Delta} \dot \pi_c \cO$.  In this case, the tri-spectrum gets its leading contribution from $\langle \cO\cO \cO \cO\rangle$ and can be written in general as
\bea
{\cal Y} =\mu^{8 - 4 \Delta} \Mp^4 |\dot H|^{2}\Big( \prod_{i=1}^4 \int  \frac{ d\tau_i  }{H^3 (- i \tau_i - \tau_0 )^3}  \langle \zeta_{\k_i}(\tau_0) \pi'_{-\k_i}(\tau_i ) \rangle' \Big)  \langle  \cO_{\k_1}(\tau_1) \cO_{\k_2} (\tau_2) \cO_{\k_3} (\tau_3) \cO_{\k_4} (\tau_4)\rangle' \
\eea
where again $\langle\dots\rangle'$ indicates dropping the momentum conserving delta function.  
Unlike the 2- and 3-point functions, the 4-point function in a CFT is not determined by symmetry in general.  However, we can say something general in the collapsed limit, where $\k_1 + \k_2 \to 0$.  As with the squeezed limit of the bispectrum, for some range of $\Delta$, we expect this will correspond to the OPE limit of the 4 point function, where $\x_1 \to \x_2$ and $\x_3 \to \x_4$.

We will take the OPE limit first, using
\beq\label{OPE}
\cO (x) \cO(0)\sim \sum_i \frac{C_i \cO_i}{x^{2\Delta - \Delta_i}} ~~~~ x\to 0,
\eeq
and then check the circumstances under which the corrections will be small.  We can write the 4-point function, in this limit as
\bea
\langle  \cO_{\k_1}(\tau_1) \cO_{\k_2} (\tau_2) \cO_{\k_3} (\tau_3) \cO_{\k_4} (\tau_4)\rangle' &\sim& \sum_{\Delta_i} \int   d^3 x_{12}d^3 x_{24} d^3 x_{34} e^{ i \k_1 \cdot \x_{12} +i (\k_1 +\k_2) \cdot \x_{24} + i \k_3 \cdot \x_{34} }\nonumber\\ && \times \frac{|C_i|^2 H^{4 \Delta}\prod_{i =1}^{4} ( i \tau_i + \tau_0)^{\Delta}}{(x_{12}^2 +\tau_{12}^2 )^{\Delta-\tfrac{ \Delta_i}{2}}(x_{34}^2 +\tau_{34}^2 )^{\Delta-\tfrac{ \Delta_i}{2}} (x_{24}^2 +\tau_{24}^2 )^{\Delta_i}} \ ,
\eea
where the sum runs over scalar operators of dimension $\Delta_i$.  We are ignoring operators with spin because the unitarity bounds imply that their dimensions will be $\geq 3$, and so, as we will see, they contribute sub-dominantly in the collapsed limit.  To make the collapsed limit clear, we made the change of variables $\x_{14} \to  \x_{12} + \x_{24}$.  In this limit, the tri-spectrum takes the form
\bea
{\cal Y} \sim \Big(\frac{\mu}{H} \Big)^{8 - 4 \Delta} \frac{\Delta_\zeta^4}{16} \frac{1}{k_1 k_2 k_3 k_4}&&\hskip -12pt  \int d\tau_1 d \tau_2 d \tau_3 d\tau_4 \Big( \prod_{i =1}^{4} (- i \tau_i -\tau_0)^{\Delta-2}(1+ i k \tau_0 \tfrac{\tau_i}{|\tau_i|} ) e^{-k_i|\tau_i|}  \Big) \nonumber \\
 \times&& \hskip -12pt \sum_{\Delta_i} \int   \frac{d^3 x_{12}d^3 x_{24} d^3 x_{34} |C_i|^2 e^{ i \k_1 \cdot \x_{12} +i (\k_1 +\k_2) \cdot \x_{24} + i \k_3 \cdot \x_{34} }}{(x_{12}^2 +\tau_{12}^2 )^{\Delta-\tfrac{ \Delta_i}{2}}(x_{34}^2 +\tau_{34}^2 )^{\Delta-\tfrac{ \Delta_i}{2}} (x_{24}^2 +\tau_{24}^2 )^{\Delta_i}} \ .
\eea
Because $k_i \gg |\k_1 + \k_2| $, it is easy to see that $|\tau_i| \lesssim k_i^{-1}$.  Therefore, $\tau_{24} \sim k_{2,4}^{-1}$ and is unimportant for the collapsed limit (notice the difference from the squeezed limit).  Therefore, the question of whether the OPE limit is applicable is determined entirely by the $x_{24}$ integral.  For $x_{24} > \tau_{24}$ this integral scales as $x_{24}^{3 - 2\Delta_i}$.  When $2 \Delta_i > 3$, this integral is dominated by the smallest values, namely $x_{24} \sim \tau_{24}$ and the OPE limit is not a good approximation.  To use the OPE to compute the collapsed limit, we must have a scalar operator of dimension $\Delta_i \leq \tfrac{3}{2}$.  

\begin{figure}[h!]
   \centering
       \includegraphics[scale =1]{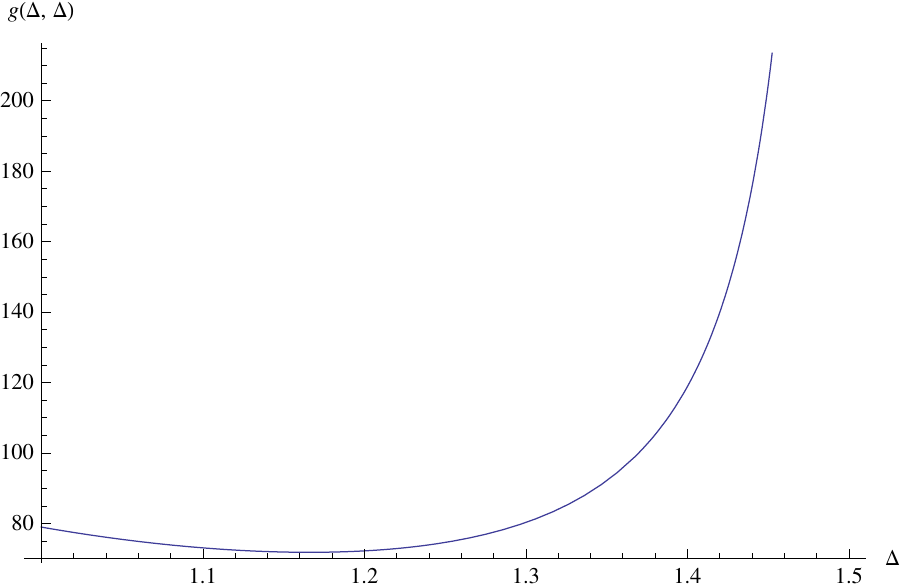}
   \caption{Numerically computed $g(\Delta, \Delta)$.}
  \label{fig:trispectrum}
\end{figure}

Following the same procedure as before, we can use equation (\ref{eq:real-2-point}) to rewrite this as
\bea
{\cal Y} &\sim&  \Big(\frac{\mu}{H} \Big)^{8 - 4 \Delta} \frac{\Delta_\zeta^4}{16} \frac{1}{k_1 k_2 k_3 k_4}  \int d\tau_1 d \tau_2 d \tau_3 d\tau_4 \Big( \prod_{i =1}^{4} (- i \tau_i -\tau_0)^{\Delta-2}(1+ i k_i \tau_0 \tfrac{\tau_i}{|\tau_i|} ) e^{-k_i|\tau_i|}  \Big) \nonumber \\
 && \times  \sum_{\Delta_i}|C_i|^2 \Big(\frac{(2\pi)^2 \Gamma(2+\tfrac{\Delta_i}{2} - \Delta)}{2^{2\Delta-\Delta_i-2} \Gamma(\Delta - \tfrac{\Delta_i}{2})}\Big)^2  (2\pi)^2 \frac{\Gamma(2-\Delta_i)}{2^{2\Delta_i-2} \Gamma(\Delta_i)} \int   \frac{d \omega_1 d\omega_2 d \omega_3}{(2\pi)^3}  \nonumber \\
 && \times    e^{ i \omega_1 \tau_{12} +i \omega_2 \tau_{24} + i \omega_3 \tau_{34}} (k_1^2 +\omega_1^2)^{\Delta - \tfrac{\Delta_i}{2} -2}(k_3^2 +\omega_3^2)^{\Delta - \tfrac{\Delta_i}{2} -2}  (|\k_1+\k_2|^2 +\omega_2^2)^{\Delta_i -2} \nonumber \ .
\eea
We notice that the above time integrals simply give factors of ${\cal J}(k, \omega, \tau_0)$ that we defined in (\ref{equ:j}).  Therefore, we can write
\bea
{\cal Y} &\sim& \Big(\frac{\mu}{H} \Big)^{8 - 4 \Delta} \frac{\Delta_\zeta^4}{k_1 k_2 k_3 k_4} \sum_i |C_i|^2 \kappa(\Delta, \Delta_i) \nonumber  \int   \frac{d \omega_1 d\omega_2 d \omega_3}{(2\pi)^3} (k_1^2 +\omega_1^2)^{\Delta - \tfrac{\Delta_i}{2} -2}(k_3^2 +\omega_3^2)^{\Delta - \tfrac{\Delta_i}{2} -2} \\ && \hskip -22pt \times    (|\k_1+\k_2|^2 +\omega_2^2)^{\Delta_i -2} {\cal J}(k_1, \omega_1,\tau_0) {\cal J}(k_2, \omega_2-\omega_1,\tau_0) {\cal J}(k_3, \omega_3,\tau_0) {\cal J}(k_4, -\omega_2 -\omega_3,\tau_0)
\eea
where we have defined
\beq
\kappa(\Delta, \Delta_i ) =\frac{1}{16} \Big(\frac{(2\pi)^2 \Gamma(2+\tfrac{\Delta_i}{2} - \Delta)}{2^{2\Delta-\Delta_i-2} \Gamma(\Delta - \tfrac{\Delta_i}{2})}\Big)^2  (2\pi)^2 \frac{\Gamma(2-\Delta_i)}{2^{2\Delta_i-2} \Gamma(\Delta_i)}  \ .
\eeq
In the collapsed limit, we have $k_1 \sim k_2$ and $k_3 \sim k_4$.  Furthermore, we see that the integral over $\omega_2$ is dominated by regions where $\omega_2 \ll \omega_1, \omega_3$ (when $\Delta_i < \tfrac{3}{2}$).  The integral over $\omega_2$ is straightforward and is given by 
\beq
\int \frac{ d \omega_2}{(2\pi) }  (|\k_1+\k_2|^2 +\omega_2^2)^{\Delta_i -2} = \frac{1}{|\k_1+\k_2|^{3 - 2 \Delta_i}} \frac{\Gamma(\tfrac{3}{2} -\Delta_i)}{2 \sqrt{\pi} \Gamma(2 - \Delta_i)} \ .
\eeq
The other integrals factorize into two copies of
\beq
\int \frac{d \omega}{(2\pi)} (k^2 +\omega^2)^{\Delta - \tfrac{\Delta_i}{2} -2}{\cal J}(k, \omega,\tau_0) {\cal J}(k, -\omega,\tau_0)\equiv \frac{4 \pi}{k^{1+ \Delta_i}} {\tilde g}(\Delta, \Delta_i) \ .
\eeq
Here we have dropped the $k$-dependence of $\tilde g(\Delta,\Delta_i)$, as it vanishes in the $\tau_0 \to 0$ limit.  Therefore, the tri-spectrum in the collapsed limit is given by
\beq
\lim_{|\k_1 +\k_2 | \to 0} \langle \zeta_{\k_1} \zeta_{\k_2} \zeta_{\k_3} \zeta_{\k_4} \rangle' = \Big(\frac{ \mu}{H}\Big)^{8 - 4 \Delta} \Delta_\zeta^4 \sum_{\Delta_i \leq \tfrac{3}{2} }\frac{(4\pi)^2|C_i|^2}{(k_1 k_3)^{3+\Delta_i} |\k_1 +\k_2|^{3-2\Delta_i} } g(\Delta,\Delta_i) \ .
\eeq
where $g(\Delta, \Delta_i)= \frac{\Gamma(\tfrac{3}{2} -\Delta_i)}{2 \sqrt{\pi} \Gamma(2 - \Delta_i)} \kappa(\Delta, \Delta_i ) \tilde g(\Delta, \Delta_i)^2$ .  
The function $g(\Delta,\Delta)$ is plotted in Fig.~\ref{fig:trispectrum}.  For a given value of $\Delta$, the dependence of $g(\Delta,\Delta_i)$ on the value of $\Delta_i$ is relatively weak until we reach the regime $\Delta_i\to 3/2$ where the OPE approximation (and hence this calculation) breaks down.  This origin of this breakdown is identical to the one we found in Section \ref{sec:squeezedbispectrum} when calculating the squeezed limit of the bispectrum. Notice that the collapsed limit is dominated by the contribution of the $\cO$'s with the smallest $\Delta_i$ that enters in the OPE. So, whenever the OPE includes operators with $\Delta_i<3/2$, we obtain the leading collapsed limit. This additionally justifies our procedure to neglect operators with spin in the OPE.

Finally, let us comment on the observability of the trispectrum.  The signal-to-noise in the trispectrum scales as $\langle \zeta^4 \rangle / \langle \zeta^2 \rangle^2$ and therefore the ratio of signal-to-noise in the trispectrum relative to the $\langle \cO \cO \cO \rangle$-bispectrum is given by
\beq
\frac{(S/N)_4}{(S/N)_3} \sim \frac{|C_i|^2}{C} \Big(\frac{\mu}{H} \Big)^{2 - \Delta} .
\eeq
For the trispectrum to give the dominant source of non-gaussianity, one requires that $|C_i|^2  \gg C$.  This arises naturally in CFTs which contain an approximate $\mathbb Z_2$ symmetry under which $\cO \to - \cO$ and $\cO_i \to \cO_i$.  When this symmetry is exact, $C= 0$ and $(S/N)_3$ vanishes (ignoring the $\langle \cO \cO \rangle$-bispectrum for the moment). This is compatible with the bounds that come from OPE associativity~\cite{Poland:2011ey}, which roughly bound $C_i$ to be smaller than order one unless $\Delta$ or $\Delta'$ are within order 10\% of 1~\footnote{A very approximate fitting formula extrapolated from the Fig.~12~of~\cite{Poland:2011ey} is $C_i\lesssim (5 \Delta'/2-3/2)(\Delta-1)$, valid for $\Delta<2$ and $\Delta'<2$.}. 

\section{Time Dependent Examples}\label{timedep}

In the conformally coupled CFT of the previous section, the unitarity bound $\Delta>1$ plays an important role, suppressing the squeezed limit of the bispectrum in favor of approximately equilateral/orthogonal non-Gaussianity.  One might naively conclude that this bound generally restricts the phenomenology arising from a CFT coupled to inflation.  However, in the presence of time dependent backgrounds, infrared physics and unitarity bounds are modified \cite{ubounds}.   Since the inflationary background is time dependent, it is important to consider this more general case.  In this section, we will see that, because of this  
scaling, the squeezed limit may take the form (\ref{BscalingCFT}) but with a larger range of powers $\Delta$, including $\Delta\approx 0$ which gives local non-Gaussianity.  To establish this it suffices to exhibit a controlled class of theories exhibiting this behavior; we will leave a more general analysis to future work.  (Non-conformal coupling of the CFT to the de Sitter curvature is a natural case for further study, for example.)        

In \cite{ubounds} the relevant effects of time-dependent couplings were analyzed in detail for a particular class of calculable examples, the renormalization group flows generated by adding a product of two single trace scalar operators ${\cal O}$ to the Lagrangian of a large-N CFT.  In the static version of this theory, these double trace flows connect a CFT in which an operator ${\cal O}_+$ has scaling dimension $\Delta_+$ to one in which this operator has been replaced by an operator ${\cal O}_-$ with scaling dimension $\Delta_-=4-\Delta_+$.  For the range of operator dimensions $1<\Delta_+<2$, the double-trace deformation by $\int {\cal O}_+^2$ is relevant, and the dimension of the scalar operator flows from $\Delta_+$ in the ultraviolet to $\Delta_-$ in the infrared.  For higher $\Delta_+$, the deformation is irrelevant and the dimension of the scalar operator becomes $\Delta_-$ in the UV; for $\Delta_+>3$ one has $\Delta_-<1$ and this introduces unitarity violation at high energies.  These statements are straightforward to derive in the large-N limit, as we will explain shortly in adapting the calculation to our application.    

As in \cite{ubounds}, we will be concerned with the generalization to the case where the double trace coupling is time dependent:  in particular, let us deform the CFT action by a power-law coupling of the form $\int \lambda t^{2\kappa} {\cal O}_+^2$ where $t=H^{-1}\log(H\tau)$.  For sufficiently large $\kappa$, specifically $\kappa> \Delta_+-2$, this renders the deformation relevant, and the theory flows to a new non-conformal theory in the infrared.  In the leading large-N limit, one can calculate the two point function of the operator to which ${\cal O}$ flows in this new theory, finding a result proportional to $1/distance^{\Delta_-}$ times powers of the time-dependent coupling.  In the next subsection, we will recover this result in our case in the process of generalizing the analysis to the inflationary background geometry.  This will give
a similar result for the two point function of the scalar operator, but now with an additional factor of $(\tau\tau')^{\Delta_-}$ reflecting the de Sitter redshifting. Because of that, up to logs we will recover the behavior (\ref{BscalingCFT}) but with $\Delta$ replaced by $\Delta_-$.  This realizes a larger range of powers of $k_L/k_S$, including the local shape for $\Delta_-=0$.   

Power-law time dependent couplings may be natural in inflation given couplings of the inflaton to other fields.
In say $m^2\varphi^2$ inflation, the inflaton rolls away from its initial value like $\phi-\phi_0\propto t$, so a linear coupling to it would naturally produce a coupling with $\kappa=1$ and so on.  For $\mu^3\varphi$ inflation we have $\varphi\propto t^{2/3}$ and one could similarly get an order-1 value of $\kappa$ if powers of $\varphi$ couple in.  That said, we will not engage in serious model-building in this work; the following scenario is just meant to establish that strongly coupled fields can introduce a wide range of scalings in the squeezed limit, including the local shape.  

\subsection{Modular example}
          
In this section, we will exhibit a concrete example which is essentially a hybrid of \cite{QSFI}\cite{Dans}\ and \cite{ubounds}.  The model includes a massive weakly interacting scalar field $\eta$  which mixes linearly both with an operator ${\cal O}$ of a large-N CFT of dimension $\Delta_+$ and with the time-derivative of the scalar perturbation $\pi_c$.  Specifically, we start from the action
\bea\label{sigmaO}
S &=& \int \frac{d\tau d\vec x}{H^4\tau^4}\left\{\frac{1}{2} \left(H^2\tau^2(\partial\eta)^2-m^2\eta^2\right) +  g_0\; (H^{-1}\log(H\tau))^\kappa \eta\, {\cal O}\right\} + S_{CFT}^{(+)} \\
 &+& \int  \frac{d\tau d\vec x}{H^4\tau^4}\left\{H\tau \rho\pi_c^\prime\eta + \mu \eta^3\right\} + S_{infl}
\eea
with $\pi_c$ the canonically normalized scalar perturbation.  
The time-dependent coupling $g(\tau)=g_0(H^{-1}\log(H\tau))^\kappa=g_0 t^\kappa$ here could come from an interaction term such as $\int (\varphi-\varphi_0)^n\eta{\cal O}$  between the inflaton $\varphi$ and the CFT sector.  As in \cite{Dans}\ the mixing term $\int\rho\dot\pi_c\eta$ could come from a coupling $\int (\partial\varphi)^2\eta$ with one factor of $\partial\varphi$  evaluated on the background and the other on the perturbation.  Finally, we have included an interaction term $\int\mu\eta^3$ which will generate a simple contribution to the non-Gaussianity.  (The CFT three-point function is down by a factor of $1/N$, so we will neglect it here but it would also be straightforward to include.)

The top line generates a time-dependent double trace flow in the strongly-coupled sector, relevant for $\kappa>\Delta_+-2$  as in \cite{ubounds}.  
In the regime where the mass term for $\eta$ dominates over its kinetic term, integrating out $\eta$ produces an operator relation 
\beq\label{etaO}
\eta = \frac{g(\tau)}{m^2}\cO 
\eeq
implying a double-trace deformation $\sim \int (g^2/2m^2)\cO^2$.  

The two-point function for $\eta$ (equivalently $\cO$) is given in the large-N limit, and to zeroth order in the cubic interaction $\mu/H$, by summing up the diagrams generated by linear mixing between $\eta$ and $\cO$.  
In flat spacetime -- applicable in our case for scales well within the horizon -- this gives  
\beq\label{etatwoflat}
\langle\eta(x)\eta(x')\rangle = \frac{-1}{c_\nu c_{-\nu} g_0^2 t^\kappa t'^{\kappa} |x-x'|^{2\Delta_-}}
\eeq
where $\nu=\Delta_+-2=2-\Delta_-$ and
\beq\label{cnu}
c_\nu = 2^{-2\nu}\pi^2 \frac{\Gamma(-\nu)}{\Gamma(2+\nu)}
\eeq
It was shown in \cite{ubounds}\ how unitarity works out in the infrared in this theory.  Using the technique in \cite{IntriligatorGrinstein}, one finds that unitarity holds -- it requires  the positivity of the product of the $-1$ in the numerator of (\ref{etatwoflat}) and a factor $(\Delta_--1)$.  The consistency of the theory at long distances  is not a surprise; it is to be expected in a theory like this which is well-defined at shorter distance scales.  
We will now derive the corresponding result in the inflationary background of interest here, which will lead to the two-point function
\beq\label{etatwodS}
\langle\eta(\tau, \vec x)\eta(\tau',{\vec x}')\rangle = \frac{-H^{2\Delta_-+2\kappa}\tau^{\Delta_-}\tau'^{\Delta_-}}{c_\nu c_{-\nu}g_0^2 (\log H\tau)^\kappa (\log H\tau')^{\kappa} \left[(\tau-\tau')^2+(\vec x-\vec x')^2\right]^{\Delta_-}}.
\eeq
This is obtained as follows.  First, we compute the correction  $\sim\int_x\int_{x'} \eta_x \langle{\cal O}_x{\cal O}_{x'}\rangle \eta_{x'}$ to the effective action in $S_{eff}$ for $\eta$ which arises through its mixing with $\cO$, giving
\beq\label{Seffeta}
S_{eff}=S_0 + \int  \frac{d\tau d\vec x}{H^4\tau^4} \int \frac{d\tau' d\vec x'}{{H^4\tau'}^4}\frac{\eta(\tau,\vec x)g_0^2(H^{-1}\log H\tau)^\kappa (H\tau)^{\Delta_+} (H^{-1}\log H\tau')^\kappa (H{\tau'})^{\Delta_+}\eta(\tau',{\vec x}') }{ \left[(\tau-\tau')^2+(\vec x-\vec x')^2\right]^{\Delta_+}}
\eeq
where $S_0$ is the original effective action for $\eta$ without the coupling to ${\cal O}$.  The trick is to note that when the double-trace term is relevant (i.e. when $\kappa>\Delta_+-2$),  the second term in $S_{eff}$ (the one generated by $\langle{\cal O}{\cal O}\rangle$) will dominate in the infrared two-point function; in \cite{ubounds}\ this was explicitly demonstrated by bounding the corrections to this approximation.  Therefore, to compute the two point function we need to just invert that term.  To do that, we can first absorb the $\tau$ and $\tau'$ dependences into a new variable
\beq\label{tildeeta}
\tilde\eta = g_0 (H^{-1} \log H\tau)^\kappa (H\tau)^{\Delta_+-4} \eta =  g_0 (H^{-1} \log H\tau)^\kappa (H\tau)^{-\Delta_-} \eta
\eeq   
The second term is then simply
\beq\label{Ssecond}
\int d\tau d\vec x d\tau' d\vec x' \frac{\tilde\eta(\tau,\vec x)\tilde\eta(\tau',\vec x')}{\left[(\tau-\tau')^2+(\vec x-\vec x')^2\right]^{\Delta_+}} 
\eeq
and we can invert it to obtain the two point function for $\tilde\eta$ exactly as in Poincare invariant Minkowski space CFT.  (Specifically, one can go to momentum space where $\langle {\cal O}{\cal O}\rangle$ is $\sim k^{2\Delta_+-4}$ and invert this to get $k^{4-2\Delta_+}=k^{2\Delta_--4}$ times appropriate constant factors \cite{ubounds}.)  This gives the usual result for double trace deformations -- a flow between $\Delta_+$ and $\Delta_-$:
\beq\label{twoetatilde}
\langle\tilde\eta \tilde\eta\rangle = \frac{-1}{c_\nu c_{-\nu}\left[(\tau-\tau')^2+(\vec x-\vec x')^2\right]^{\Delta_-}}
\eeq
Finally, putting back the $\tau$ dependence from (\ref{tildeeta}), we obtain the claimed result (\ref{etatwodS}). 

The correction we generated to the power spectrum in this theory behaves paremetrically as
\beq\label{modularpower}
\Delta{\cal P}_\zeta \sim {\cal P}_{\zeta} \left(\frac{\rho}{H}\right)^2 \frac{H^{2\Delta_-+2\kappa-2}}{\hat g_0^2}\frac{1}{\log^\kappa(H/k)\log^\kappa(H/k)}
\eeq
As we will see, the non-Gaussianity is determined by similar factors, but is enhanced by a factor of $\Delta_\zeta^{-1}\sim 10^5$.  We will focus on the case where $\Delta{\cal P}_\zeta < {\cal P}_\zeta$, for which the leading effect of the coupling to our strongly coupled sector is in the non-Gaussianity.   

As already mentioned, we will take  the $\int\mu\eta^3$ term as the leading source of the non-Gaussianity.  This leads to a calculation very similar to that arising in the case of additional weakly coupled fields \cite{QSFI}\cite{Dans}, and we will compare the two as we go.  Schematically the leading contribution to the bispectrum is 
\beq\label{threeschem}
B(k_1,k_2,k_3) \sim   \int \frac{d\tau}{H^4\tau^4}\mu\prod_{j=1}^3\left\{\int\frac{d\tau_j }{H^3\tau_j^3} \rho \, G_\eta(k_j;\tau_j,\tau) \partial_{\tau_j}G_{\pi_c\zeta}(k_j;\tau_j,0) \, \right\}
\eeq   
Here $G_{\pi_c\zeta}=G_\zeta M_P\sqrt{2|\dot H|}/H= G_\zeta H/\Delta_\zeta$ is proportional to the two-point Green's function of the scalar perturbation $\zeta$, and
\beq\label{Gzeta}
\partial_{\tau_j} G_\zeta(k_j;\tau_j,0)=\partial_{\tau_j}\left\{\frac{\Delta_\zeta^2}{k_j^3} \,(1+ k_j|\tau_j|)e^{-k_j|\tau_j|}\right\}=-\Delta_\zeta^2\,\frac{\tau_j}{k_j}\, e^{-k_j|\tau_j|}
\eeq
(working in Euclidean signature).
In (\ref{threeschem}),  $G_\eta$ is the two-point Green's function of $\eta$, given by  Fourier transforming (\ref{etatwodS}):
\beq\label{twoetak}
G_\eta(k_j;\tau,\tau_j)  =\frac{ 2^{5/2}\pi^{3/2}}{c_\nu c_{-\nu}}\frac{H^{2\Delta_-+2\kappa}}{\Gamma(\Delta_-)g_0^2}\frac{k_j^{2\Delta_--3}(\tau\tau_j)^{\Delta_-}K_{3/2-\Delta_-}(k_j|\tau-\tau_j|)}{|k_j(\tau-\tau_j)|^{\Delta_--3/2}\log^\kappa(H\tau)\log^\kappa(H\tau_j)}
\eeq

Before continuing, let us remark on the combination 
\beq\label{hatg}
\hat g_0\equiv \sqrt{\Gamma(\Delta_-)}g_0
\eeq
 which appears here since we will be particularly interested in the regime $\Delta_-\to 0$, and the $\Gamma$ function has a pole there.  This effect is also evident in the position-space propagator, where 
$1/|x-x'|^{2\Delta_-}=1-2\Delta_-\log |x-x'|+\dots$ as $\Delta_-\to 0$;  the two point function       
of $\eta$ only depends on $\x-\x'$ through the log piece proportional to $\Delta_-$.  The term independent of $\x-\x'$ will not contribute to our three point function at finite spatial momentum $\k$, so the leading term we need in the two point function of $\eta$ is the log term proportional to $\Delta_-$.
If we take $\hat g_0$ to be finite as $\Delta_-\to 0$, then the two-point function of $\eta$, including the effects of its mixing with $\cO$, stays finite.  It is this degree of freedom which couples linearly to the inflationary perturbation, and it seems natural to keep its propagator of order 1.    
This regime is consistent with the flow we are working with \cite{ubounds}\ occurring within the horizon.  To see this, one can consult equations C.1-C.3 of \cite{ubounds}, whose leading $x-x'$ dependent term depends on $g_0$ and $\Delta_-$ through the combination $\hat g_0$.         

From these expressions (\ref{Gzeta}) and (\ref{twoetak}), we can now determine the leading contribution to the integrals in (\ref{threeschem}) in the squeezed limit $k_2\sim k_3 \gg k_1 $.   For the short modes (\ref{Gzeta}) exponentially suppresses the integral for $\tau_{2,3}>1/k_2$.  As a result, the Bessel function  $K_{\Delta_--3/2}(k|\tau-\tau_j|)$ suppresses the integral for $\tau>1/k_2$.       

Next, consider the opposite limit of small $k\tau$ in our integral.  
The two-point function (\ref{twoetak}) is different from the two-point function of massive fields, as is easy to see in position space where the latter is a hypergeometric function whereas ours is the simpler function (\ref{etatwodS}).  However, they behave similarly at small $k\tau$ and $k\tau'$, where for $\Delta_-<3/2$ (\ref{twoetak}) becomes
\beq\label{smallarg}
\sim \frac{H^{2\Delta_-+2\kappa}\Gamma(3/2-\Delta_-)}{ k_j^3 c_\nu c_{-\nu}}\frac{(k_j\tau)^{\Delta_-}(k_j\tau_j)^{\Delta_-}}{{\hat g_0}^2 \log^\kappa(H\tau)\log^\kappa(H\tau_j)}
\eeq      
Up to the logs, this is just like the corresponding expression for weakly coupled fields of mass $\tilde m$ (quasi-single-field inflation), with the identification
\beq\label{massDelta}
\Delta_- \leftrightarrow 3/2-\sqrt{9/4-{\tilde m}^2/H^2}
\eeq
The $\tau_j$ integrals in (\ref{threeschem}) are manifestly dominated by the largest value $\tau_j\sim 1/k_j$ allowed by the exponentials.  As in \cite{QSFI}\ and the example of the previous section, the small $\tau$ behavior does not lead to large effects; again, this can be seen from the contour prescription \cite{Behbahani:2012be}.

Given that, the integral is well approximated by a saddle point with $\tau\sim \tau_{2,3}\sim k_S^{-1}$ and $\tau_1\sim k_L^{-1}$.  Plugging this into (\ref{threeschem}) it is easy to read off the scaling and amplitude in the squeezed limit
\bea
B(k_1,k_2,k_3) &\sim& \frac{ f_{NL}\, \Delta_\zeta^4 }{k_L^3 k_S^3}\left(\frac{k_L}{k_S}\right)^{\Delta_-}\\
&{\rm with}&  \label{OetafNL} \nonumber \\  \nonumber
  f_{NL} &\sim& \left(\frac{\rho}{H}\right)^3\left(\frac{\mu}{H}\right)\Delta_\zeta^{-1} \left(\frac{H^{\Delta_-+\kappa -1}}{\hat g_0}\right)^6 \\ 
\eea
up to logarithmic factors.  Here we have dropped factors that are order 1 in the regime $\Delta_-\ll 1$ which approaches the local shape.

\begin{figure}[h!]
   \centering
       \includegraphics[scale =0.4]{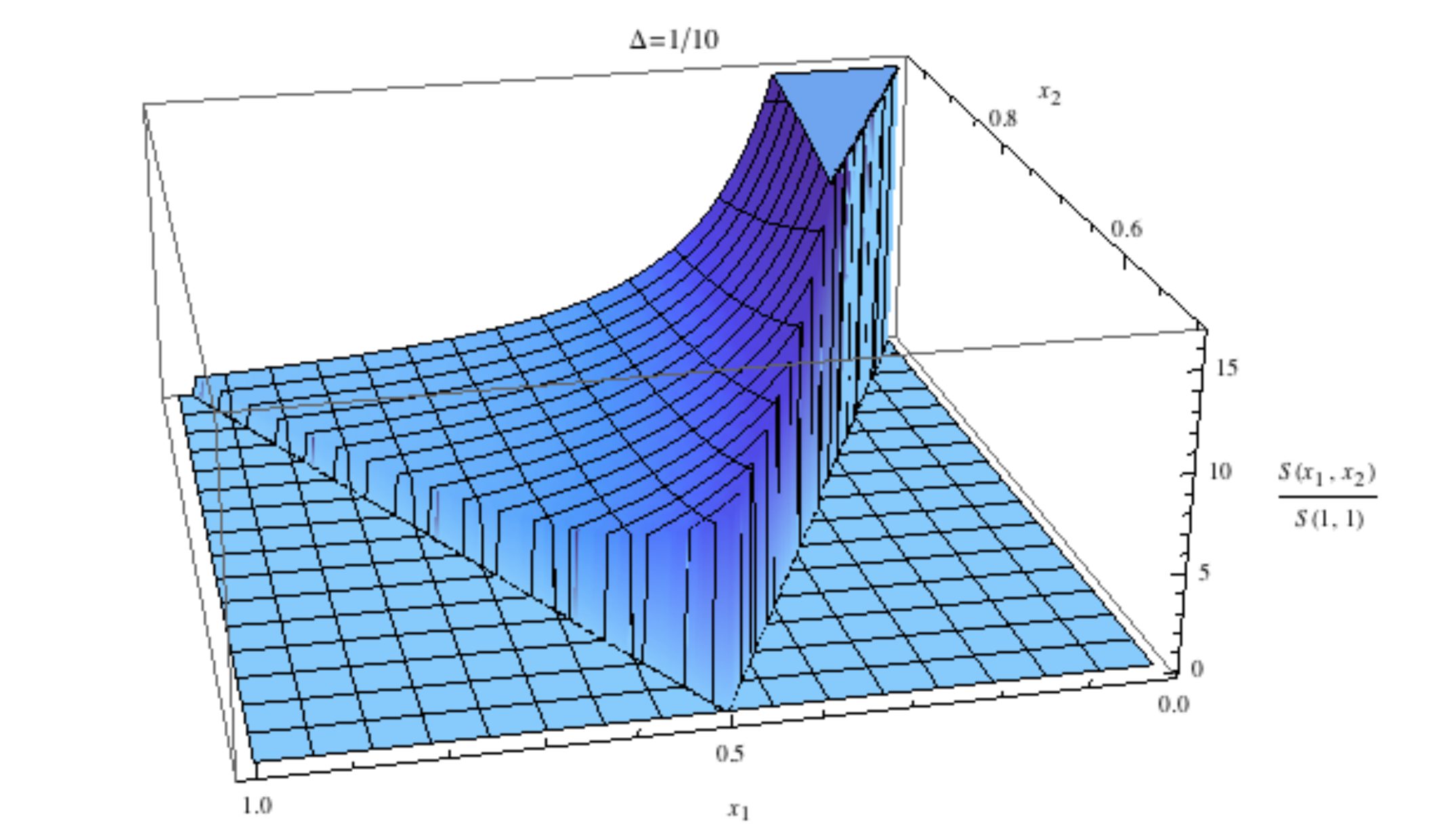}
   \caption{Numerically computed shape function, $S(x_1,x_2)$ evaluated for $\Delta_- = \frac{1}{10}$ and $\kappa =1$.}
  \label{fig:shapeS3}
\end{figure}

The shape can be computed numerically starting from (\ref{threeschem}).  The case of $\Delta_- = \frac{1}{10}$ and $\kappa =1$ is shown in Fig.~\ref{fig:shapeS3}.  The bispectrum is not scale invariant with an effective $f_{\rm NL}$ that grows logarithmically at larger scales (smaller $k$).  As a result,  shape function is enhanced in the squeezed limit by additional factors of $1/ \log x_2$, which represent the stronger interaction at large scales. Numerically, we find significant overlap with the local shape for all $\Delta_- \leq 1$ ($\cos(S_{\Delta_-},S_{\rm local}) \gtrsim 0.8$).  As discussed around Fig.~\ref{fig:fsqueezed}, the numerical error appears to grow as we approach the squeezed limit, which is also where the signal-to-noise is dominated.  For this reason, the specific values of the cosine with the local shape are likely not meaningful.  
\vskip 8pt
Let us compare our amplitude to the case of an interaction $\int\dot\pi_c\sigma$ between the inflaton perturbation and a weakly interacting field $\sigma$ of mass $\tilde m$. 
In the weakly coupled case, the local shape arises for massless scalars, $\tilde m\to 0$.  But the cubic interaction -- which here produces the non-Gaussianity -- would generate a mass $\tilde m\sim \mu$, unless it is tuned away.  Without such tuning, the amplitude is proportional to the mass through its dependence on $\mu$.    
In our case, although we have the $\mu\eta^3$ coupling, we do not need to take the mass $m$ of $\eta$ to zero; instead we get the local shape from $\Delta_-\ll 1$ rather than $m\ll H$.   Our amplitude $f_{NL}$ can easily be large for small $\Delta_-$, with the flow we have used fitting well inside the Hubble horizon. 
Finally, we also note that the calculation of the two point function in this example requires a large-N limit to control as in \cite{ubounds}, which may be regarded as a small (perhaps ten percent) tune we have introduced for calculational convenience.        

As a final comment on the amplitude of the non-Gaussianity, let us comment briefly on the sensitivity this gives to higher dimension operators in the theory (\ref{sigmaO}).  This example is more natural than the simple case (\ref{dimsix}) discussed in the introduction in that the latter involved a relevant operator, whereas here we may consider a nearly marginal operator $\cO_{\Delta_+\approx 4}$ in the UV, with a relevant flow induced by the time dependent background.  Let us make a quick estimate of the precision measurements available via $f_{NL}$ in this example.  
Focusing on the case $\Delta_+\approx 4, \Delta_-\approx 0$, $\kappa\approx 2$ and assuming a common scale $\tilde M_*$ suppressing the higher dimension operators in the UV theory, we have couplings of the form 
\beq\label{higherdimthree}
\int \left\{ \frac{(\phi-\phi_0)^2\eta \cO}{\tilde M_*^3}+\frac{(\partial\phi)^2\eta}{\tilde M_*}+\dots\right\}
\eeq
In terms of the parameters used above, we have 
\beq\label{newparam}
\hat g_0\sim \frac{\dot\phi^2}{\tilde M_*^3}\sim \frac{H^4\Delta_\zeta^{-2}}{\tilde M_*^3} ~~~~~{\rm and}~~~~~\rho \sim \frac{\dot\phi}{\tilde M_*}\sim \frac{H^2\Delta_\zeta^{-1}}{\tilde M_*}   
\eeq
where we used the slow-roll relation $\Delta_\zeta\sim H^2/\dot\phi$.  
In order for the double-trace flow to fit within our Hubble patch, we need $\hat g_0 =H/\epsilon_0$ with $\epsilon_0<1$.  This implies
\beq\label{HM}
\frac{H^3}{\tilde M_*^3}\sim \frac{\Delta_\zeta^2}{\epsilon_0}
\eeq      
with
\beq\label{fNLM}
f_{NL}\sim \Delta_\zeta^{-4}\frac{H^3}{\tilde M_*^3}\left(\frac{\mu}{H}\right)\epsilon_0^6\sim \Delta_\zeta^{-2}\left(\frac{\mu}{H}\right)\epsilon_0^5
\eeq
To get a rough estimate of the sensitivity here, a null result $f_{NL}\lesssim 1$ would probe the regime $\mu\sim H$ and $\epsilon_0\sim 10^{-2}$ with $\tilde M\sim 10^{8/3}H$.

\section{Discussion and Future Directions}

In this work we have analyzed the non-Gaussian corrections to the inflationary perturbations arising from a linear coupling to a scalar operator ${\cal O}$ in a strongly coupled theory.  Specifically, we analyzed two simple, calculable cases: a conformally coupled CFT (at least near the Hubble scale), and a particular type of time-dependent deformation thereof.  We found characteristic scaling behavior of the bispectrum near the squeezed limit, going like  $\sim (k_L/k_S)^\gamma$ times the local shape.  This behavior is similar to that generated by weakly coupled massive fields \cite{QSFI}\cite{Dans},  but with the exponent $\gamma$ in our case depending on the dimension of ${\cal O}$.  Our two examples together exhibit a large range of exponents $\gamma$, and a range of shapes, including equilateral and orthogonal for large $\gamma$ and the local shape for $\gamma\to 0$.  In the nearly-equilateral case, it is at least partially degenerate observationally with single field inflation (e.g. \cite{DBI,generalsingle,EFT,Senatore:2009gt}).  As in that case,
the amplitude can naturally be large and gives us an observational probe of higher dimension operators.  In particular, we have seen here that non-Gaussianity provides a precision test for additional sectors of fields coupled through higher-dimension operators suppressed by a scale $M_*\gg H$, and either a detection or a null result would be very informative.

There are many interesting generalizations.  Here we considered perturbative couplings between the inflationary and strongy coupled field theory sectors, leading to small corrections to the power spectrum along with relatively substantial contributions to the non-Gaussianity.  It would be interesting to consider stronger mixing interactions, requiring resummation to determine the power spectrum (something which simplifies somewhat if one considers a large N limit).   
Here also we focused on scalar operators, but higher spin operators could lead to their own distinctive effects, generalizing \cite{nicolis}\ to the case of strongly coupled fields.   Another interesting direction is to understand more systematically how non-conformal couplings affect the bispectrum in interacting theories, a question amenable to perturbative quantum field theory calculations of anomalous dimensions in some interesting limits.  

In another direction, UV complete mechanisms for inflation studied thus far often involve strongly coupled sectors which play a key role in producing dynamically the small scales required in the inflaton effective action.  It will be interesting to determine the implied couplings between their operators ${\cal O}$ and the inflaton perturbations in such examples, a potential source of new signatures or constraints.  Related to this, it would be worthwhile to assess more systematically the level of Wilsonian naturalness in various models of multifield perturbations, in preparation for the observational results which will determine its actual natural-ness.

\section*{Acknowledgements}

We thank Masha Baryakthar for initial collaboration in the project. We thank I. Antoniadias, D. Baumann, X. Dong, B. Horn, K. Smith, G. Torroba, L. Verde and S. Rychkov  for helpful discussions.  This work was supported in part by the National Science Foundation under grant PHY-0756174 and by the Department of Energy under contracts DE-AC03-76SF00515 and DE-FG02-92-ER40699. L.S. is supported by DOE Early Career Award DE-FG02-12ER41854 and the National Science Foundation under PHY-1068380. M.Z. is supported by the National Science Foundation under PHY-0855425, AST-0907969 and PHY-1213563 and by the David and Lucile Packard Foundation.

\bigskip
\appendix

\section{Radiative Corrections}\label{app:tuning}

In this appendix we discuss radiative corrections to the effective action of the models considered in Section \ref{sec:conformal}.  As discussed in the main text, 
the relative sizes of the couplings in (\ref{equ:doublemix}) determine which correlation functions of our operators $\cO$ generate the dominant contribution to the bispectrum.  Because the size of the couplings is also important for radiative corrections, we must evaluate the radiative stability for the different cases separately, and our analysis in this appendix will follow the order we take in the main text.  In all cases, we will ask what is the natural value of $f_{\rm NL}$ from the tree level Hamiltonian and compare to those generated by loops. 

\subsection{Case 1: Bispectrum determined by $\langle \cO \cO \rangle$}
As in Section \ref{sec:boo}, we will consider the interaction Hamiltonian 
\beq\label{equ:doublemix}
 {\cal H}_{\rm int} =\tfrac{1}{2}  \mu^{2 - \Delta} \Mp 
|\dot H|^{1/2} ( 2 \dot \pi - \partial_\mu \pi \partial^\mu \pi) \cO + \tfrac{1}{4} \Mp^2 |\dot H| \tilde \mu^{- \Delta} ( - 2 \dot \pi + \partial_\mu \pi \partial^\mu \pi)^2 \cO  \ .
 \eeq
Contribution of the second operator gives the dominant contribution to the bispectrum when $\tilde \mu^\Delta\lesssim\mu^{2-\Delta}(\Mp\dot H)^{1/2}$, which is the regime of interest in this subsection.  Because the second operator is always irrelevant, the interaction becomes strongly coupled (i.e. perturbative unitarity breaks down) at the scale
 \beq
 \Lambda_U\sim \tilde \mu \ .
 \eeq
 Another scale in the problem is $f_\pi^2\sim M_P|\dot H|^{1/2}$, the time translation symmetry breaking scale.    
 
 Let us further assume $\langle \cO^3 \rangle$ is negligible, so that the dominant contribution to $f_{\rm NL}$ is from
\beq
\langle \zeta^3 \rangle \sim  \langle \zeta \dot \pi_c \rangle^3 \, \mu^{2-  \Delta}\tilde \mu^{-\Delta} \langle \cO^2 \rangle  
\eeq
which gives
\beq
f_{\rm NL}^{(1)} \sim \Big(\frac{\mu }{H}\Big)^{2 - \Delta}\Big(\frac{H}{\tilde \mu} \Big)^{\Delta} \Delta_\zeta^{-1} \ .
\eeq
Clearly this can be large. Let us first consider the renormalization of the term in the Lagrangian proportional to  $\cO$.  This is generated by a loop of $\pi$'s to give 
\beq
\H_{\rm int}^{\rm rad.} \supset \frac{\Lambda^4}{\tilde\mu^\Delta}\cO\ .
\eeq
This is a relevant operator only for $\Delta<4$. In this case, we have to impose that
\beq
\frac{\Lambda^4}{\tilde\mu^\Delta}\lesssim H^{4-\Delta}\quad\Rightarrow\quad \Lambda\lesssim H \left(\frac{\tilde\mu}{H}\right)^{\Delta/4}\ .
\eeq
This can be clearly satisfied for $\tilde\mu\gg H$ while at the same time having $\Lambda\gg H$. However, if we push $\Lambda$ to be as high as the $\Lambda_U$, we find that the condition becomes $\tilde\mu< H$, which cannot be accepted. This means that in order for having large effects from this second operator, we must have $\Lambda\ll \Lambda_U$.  This requires an additional scale beyond $\Lambda_U$ to arise in the UV completion of our model.  One possibility is that that role is played by $f^2_\pi\sim \Mp|\dot H|^{1/2}$, which is also a physical scale in the system.

Another term which is generated by renormalization is
\beq\label{equ:radiative3}
\H_{\rm int}^{\rm rad.} \supset \Big(\frac{\mu}{\Lambda}\Big)^{2 - \Delta}   \frac{(\Mp^2 |\dot H|)^{3/2} }{\Lambda^{2-\Delta} \tilde \mu^{\Delta} }( - 2 \dot \pi + \partial_\mu \pi \partial^\mu \pi)^3 \ ,
\eeq
which generates a contribution to $f_{\rm NL}$ of order
\beq
f_{\rm NL}^{(2)} \sim \Big(\frac{\mu}{\Lambda}\Big)^{2 - \Delta} \Big(\frac{\Lambda}{\tilde \mu} \Big)^{\Delta} \frac{\Mp |\dot H|^{1/2}}{\Lambda^2} \sim f_{\rm NL}^{(1)} \Big(\frac{H}{\Lambda} \Big)^{4-2 \Delta} \ .
\eeq
For $\Delta> 2$ we have $f_{\rm NL}^{(2)} \gg f_{\rm NL}^{(1)}$, which tells us that in the case $\Delta>2$ non-gaussianities from the second operator in~(\ref{equ:doublemix}) are subdominant to those induced by radiative corrections, so they do not dominate in technically natural theories. 

This situation will not extend to the case in the 
next subsection.  There we find theories with $\Delta<8/3$ for which the theory is technically natural and the leading signal is from the three-point function of the conformal sector, provided the latter has sufficiently large amplitude $C$.  For $C\simeq 0$, we will find technically natural theories with large non-Gaussianities only for $\Delta<2$, and these will require a crossover to a UV completion at a scale below $\Lambda_U$.

\subsection{Case 2: Bispectrum determined by $\langle \cO \cO \cO \rangle$}
 As in Section \ref{sec:booo}, when $\langle \cO \cO \cO \rangle$ gives the dominant contribution to the bispectrum, we only require the interaction
 \beq\label{equ:mixing}
 \H_{\rm int} = \mu^{2 - \Delta} \Mp 
|\dot H|^{1/2} ( - 2 \dot \pi + \partial_\mu \pi \partial^\mu \pi) \cO \ .
 \eeq
For $\Delta \leq 2$, the leading operator is relevant, and therefore, for control we require $\mu < H$.  On the other hand, when $\Delta >2$, the leading term is irrelevant and therefore $\mu > H$ is required for control.

\subsubsection{Relevant deformation: $\Delta\leq 2$}

When $\Delta \leq 2$ only the second operator in (\ref{equ:mixing}) is irrelevant.  The strong coupling scale associated to this interaction is therefore set by
\beq\label{eq:unitarity}
\Lambda_U\sim \frac{(\Mp^2|\dot H|)^{\frac{1}{2\Delta}}}{\mu^{\frac{2-\Delta}{\Delta}}}\ .
\eeq
Because $\mu < H$, we see that $\Lambda^2_U \geq f^2_{\pi} \sim \Mp |\dot H|^{1/2}$.

Schematically, the three-point function receives two tree level contributions:
\beq
\langle \zeta^3 \rangle \sim  \langle \zeta \dot \pi_c \rangle^3 \Big[\frac{ \mu^{6- 3 \Delta}}{H^3}\langle \cO^3 \rangle + \frac{\mu^{4- 2 \Delta}}{ \Mp |\dot H|^{1/2}\;H^2} \langle \cO^2 \rangle  \Big]\ .
\eeq
Assuming the absence of UV divergences when we take these expectation values for $\Delta<2$ (a fact we will recover shortly)  
the the natural values of $f_{\rm NL}$ are given by
\bea
f_{\rm NL}^{(1)} &\sim&  C   \Big(  \frac{ \mu}{H}\Big)^{6- 3 \Delta}  \Delta_\zeta^{-1}   \ ,        \\ \label{eq:mixed_contribution1}
f_{\rm NL}^{(2)} &\sim&     \Big(  \frac{ \mu}{H}\Big)^{4- 2 \Delta} \frac{H^2}{\Mp |\dot H|^{1/2} }  \Delta_\zeta^{-1}  \sim     \Big(  \frac{ \mu}{H}\Big)^{4- 2 \Delta} \lesssim 1 \ ,
\eea
where $C$ is the coefficient of $\langle \cO^3 \rangle $.  Here we see that only $f_{\rm NL}^{(1)}$ has the possibility of being large.

Now let us look at which operators are generated by loops or by CFT dynamics, given a hard cutoff at the scale $\Lambda$. We must have $H\ll \Lambda\leq \Lambda_U$. 
Schematically, we expect to generate the following operators
\bea\label{equ:radiative}
\H_{\rm int}^{\rm rad.} &=& \mu^{2 - \Delta} \frac{\Lambda^4}{\Mp |\dot H|^{1/2} }\cO + C \frac{\mu^{4 - 2 \Delta}\Mp^2| \dot H|}{\Lambda^{4 - \Delta} } (- 2 \dot \pi + \partial_\mu \pi \partial^\mu \pi)^2 \cO \\ &&+\Mp^2 |\dot H| \frac{\mu^{4-2 \Delta}}{\Lambda^{4-2 \Delta}} (- 2 \dot \pi + \partial_\mu \pi \partial^\mu \pi)^2  +C \frac{(\Mp^2 |\dot H|)^{3/2} \mu^{6-3\Delta}}{\Lambda^{8 - 3 \Delta}} (- 2 \dot \pi + \partial_\mu \pi \partial^\mu \pi)^3 \nonumber \ .
 \eea
The most relevant term is the first one and is dangerous because it can break the conformal symmetry. In order for this term to be negligible, the effective scale of the breaking must be lower than Hubble.  This can be achieved, provided that
\beq
 \frac{\Lambda^4}{\Lambda_U^{\Delta} }
 \ll H^{4- \Delta} \quad\Rightarrow\quad \Lambda\ll H \left(\frac{\Lambda_U}{H}\right)^{\Delta/4} \ .
\eeq
This can be satisfied for $H\ll \Lambda\ll \Lambda_U $. However, we cannot push $\Lambda$ all the way up to $\Lambda_U$ because in this case the constraint would become $\Lambda_U\ll H$, which is not acceptable. We have not explicitly UV completed our theory; it may happen in such a completion that additional operators appear suppressed by the lower scale $f_\pi\sim \Mp |\dot H|^{1/2}$, or at some other scale $\Lambda$ that depends on the dynamics of additional sectors.  We will not engage in serious model building here,\footnote{something to potentially return to depending on the outcome of measurements of $f_{\rm NL}$.}, but note that our low energy model as written only pertains below the scale $\Lambda\ll \Lambda_U$.      

By looking at the quadratic term $\dot\pi^2$, we see that a speed of sound slightly different from one is generated
\beq
\Delta_\zeta^{1/4}\left(\frac{\mu}{H}\right)^{(6-\Delta)/4}\lesssim \frac{\delta c_s^2}{c_s^2}\sim \frac{\mu^{4-2 \Delta}}{\Lambda^{4-2 \Delta}} \ll 1 \ .
\eeq
Now let us look at the contributions to $f_{\rm NL}$ from the second through fourth terms in (\ref{equ:radiative}) :
\bea
f_{\rm NL}^{(3)} &\sim&   \   C\Big(  \frac{ \mu}{H}\Big)^{6- 3 \Delta}  \Big(  \frac{ H}{\Lambda}\Big)^{4- \Delta} \Delta_\zeta^{-1} \ll f_{\rm NL}^{(1)} \ , \\
f_{\rm NL}^{(4)} &\sim&  \frac{1}{c_s^2}(1-c_s^2) \sim \frac{\mu^{4-2 \Delta}}{\Lambda^{4-2 \Delta}} \ll 1  \ ,\\
f_{\rm NL}^{(5)} &\sim& \frac{\Mp |\dot H|^{1/2}}{\Lambda^2} \frac{\mu^{6-3 \Delta}}{\Lambda^{6- 3\Delta }} \ll \frac{\Mp |\dot H|^{1/2}}{H^2} \frac{\mu^{6-3 \Delta}}{H^{6 - 3\Delta}} \sim f_{\rm NL}^{(1)} \ .
\eea
We see that $f_{\rm NL}^{(1)}$ gives the dominant contribution to the bispectrum. This means that for $\Delta<2$ the theory is technically natural.

\subsubsection{Irrelevant Deformation : $\Delta > 2$}

To distinguish the two cases, we will redefine $\mu \to f$.
Let us identify the strong coupling scale. Since there are two irrelevant operators, the unitary bound is given by the smallest scale suppressing these two operators. We have
\beq
\Lambda_U={\rm Min}[f, f^{1-2/\Delta} (\Mp^2\dot H)^{1/(2\Delta)}]\equiv{\rm Min}[\Lambda_{U,1},\Lambda_{U,2}]\ .
\eeq
Notice that since we must have $\Lambda_U\gg H$, this implies $f\gg H$.
Assuming no UV divergences, there are again two contributions to the bispectrum with 
\bea
f_{\rm NL}^{(1)} &\sim&  C   \Big(  \frac{ H}{f}\Big)^{ 3 \Delta - 6 }  \Delta_\zeta^{-1}    \ ,      \\ \label{eq:mixed_contribution2}
f_{\rm NL}^{(2)} &\sim&     \Big(  \frac{ H}{f}\Big)^{2 \Delta-4} \frac{H^2}{\Mp |\dot H|^{1/2} }  \Delta_\zeta^{-1}  \sim     \Big(  \frac{ H}{f}\Big)^{2 \Delta- 4} \lesssim 1 \ .
\eea
The similarity to the first case is expected here because nothing has changed, at this level.  The differences will be more apparent when we discuss radiative corrections.

We will again look at renormalization of various operators assuming a hard cutoff $\Lambda$, with $H\ll \Lambda\leq \Lambda_U$.  As expected, we will renormalize the same operators as the previous section:
\bea\label{equ:radiative2}
\H_{\rm int}^{\rm rad.}& =& \frac{1}{f^{\Delta -2}} \frac{\Lambda^4}{\Mp |\dot H|^{1/2} }\cO + C \frac{\Mp^2| \dot H|}{f^{2\Delta-4}\Lambda^{4 - \Delta} } (- 2 \dot \pi + \partial_\mu \pi \partial^\mu \pi)^2 \cO \\ 
&&+\Mp^2 |\dot H| \frac{\Lambda^{2 \Delta-4}}{f^{2\Delta -4}} (- 2 \dot \pi + \partial_\mu \pi \partial^\mu \pi)^2+ C \frac{(\Mp^2 |\dot H|)^{3/2} }{f^{3 \Delta - 6} \Lambda^{8 - 3 \Delta}} (- 2 \dot \pi + \partial_\mu \pi \partial^\mu \pi)^3
\eea
For $\Delta > 4$, the first term is irrelevant to does not introduce a constraint.  However, for $\Delta <4$, we require
\beq
 \frac{\Lambda^4}{\Lambda_{U,2}^\Delta } \ll H^{4- \Delta} \quad\Rightarrow\quad \Lambda < H\left(\frac{\Lambda_{U,2}}{H}\right)^{\Delta/4}\ .
\eeq
In the case in which the unitarity bound is given by $\Lambda_{U,2}$ (that is for $f\gtrsim (\dot H\Mp^2)^{1/4}\sim H \Delta_\zeta^{-1/2}$), $\Lambda$ cannot be pushed all the way to $\Lambda_{U,2}$, as this constraint would require $\Lambda_{U,2}\ll H$.  However, when the unitarity bound is  $\Lambda_{U,1}$ (that is when $f\lesssim H \Delta_\zeta^{-1/2}$), it is possible to push $\Lambda$ to $\Lambda_{U,1}$ by imposing the stronger constraint $H\ll f\ll  H \Delta_\zeta^{-1/(6-\Delta)}$.  This feature of the model, that is the fact that we do not induce the most relevant operator even in the case where the cutoff is as high as the unitarity bound, can be understood in terms of an approximate $\pi\to-\pi,\; \cO\to-\cO$ $Z_2$ symmetry that is softly broken by the operator $(\partial\pi)^2\cO$. A very similar $Z_2$ approximate symmetry is what allows to have a leading trispectrum in single field inflation~\cite{Senatore:2010jy}.

The correction to the power spectrum scales as $(\Lambda/f)^{2\Delta-4}$, which is small if $\Lambda\lesssim f$, which is always the case if $f\lesssim (\dot H\Mp^2)^{1/4}$. A speed of sound different from one is generated. We also require $\Lambda > H$, which is always consistent with this constraint.  Again, let us look at the contributions to $f_{\rm NL}$ from the second to fourth term in (\ref{equ:radiative2}) :
\bea
f_{\rm NL}^{(3)} &\sim&   \   \Big(  \frac{ H}{f}\Big)^{ 3 \Delta-6 }  \Big(  \frac{ H}{\Lambda}\Big)^{4- \Delta} \Delta_\zeta^{-1}  \ , \\
f_{\rm NL}^{(4)} &\sim&  \frac{1}{c_s^2}(1-c_s^2) \sim \frac{\Lambda^{2 \Delta-4}}{f^{2\Delta -4}} \lesssim 1 \ , \\
f_{\rm NL}^{(5)} &\sim& \frac{\Mp |\dot H|^{1/2}}{\Lambda^2} \frac{\Lambda^{3 \Delta-6}}{f^{3\Delta -6}} \sim \frac{\Mp |\dot H|^{1/2}}{H^2} \frac{H^{3 \Delta-6}}{f^{6 \Delta-6}} \Big(\frac{\Lambda}{H} \Big)^{3 \Delta -8} \ .  
\eea
For $\Delta > \frac{8}{3}$ we have $f_{\rm NL}^{(5)} \gg f_{\rm NL}^{(1)}$.  For $\Delta > 4$, $f_{\rm NL}^{(3)} \gg f_{\rm NL}^{(1)}$.  This means that in the case $\Delta>8/3$ the signature from the three-point function of conformal operators is not  the leading one for technically natural theories.

\subsection{Case 3: Trispectrum determined by $\langle \cO \cO \rangle$}

If we start with the Lagrangian (\ref{equ:doublemix}) with $\mu=0$, we have that the leading signal in the trispectrum induced by $\langle\cO\cO\rangle$ with a size $\tau_{\rm NL} \Delta_\zeta^2\sim(\tilde\mu/H)^{-2\Delta} $. Naively, the operator has a $Z_2$ symmetry under which $\pi\to-\pi$ that forbids the appearance of any odd power of $\pi$. This naively forbids the presence of operators that would induce a non-vanishing bisectrum. However, the non-linear realization of time-diffs. imposes that the operator $(\delta g^{00})^2\cO\supset \dot\pi^2\cO$ also contains cubic terms in~$\pi$. Under a loop, these terms will generate terms of the form $\dot\pi\cO$. If $\Lambda\leq \Lambda_U\sim\tilde\mu$ is the cutoff of the loops, we have
\beq
\H_{\rm int}^{\rm rad.}\supset \mu_{\rm rad.}^{2-\Delta}(\dot H\Mp^2)^{1/2}\dot\pi\cO\ , \qquad \mu_{\rm rad.}^{2-\Delta}\sim\frac{\Lambda^4\Delta_\zeta}{H^2\tilde\mu^\Delta}\ .
\eeq
Let us see under which conditions on the cutoff the induced bispectrum is smaller than the trispectrum. We have
\bea
&&f_{\rm NL}^{\langle\cO\cO\rangle}\sim \frac{1}{\Delta_\zeta}\left(\frac{\tilde\mu}{H}\right)^{\Delta}\left(\frac{\mu_{\rm rad.}}{H}\right)^{2-\Delta}\sim\frac{\Lambda^4}{H^4}\left(\frac{H}{\tilde\mu}\right)^{2\Delta}\ ,\\ \nonumber
&&f_{\rm NL}^{\langle\cO\cO\cO\rangle}\sim \frac{C}{\Delta_\zeta}\left(\frac{\mu_{\rm rad.}}{H}\right)^{3(2-\Delta)}\sim\frac{C}{\Delta_\zeta^2}\frac{\Lambda^{12}}{H^{12}}\left(\frac{H}{\tilde\mu}\right)^{3\Delta}\ .
\eea
Imposing the signal from both to be smaller than the one on the trispectrum implies respectively:
\bea
\Lambda^2\lesssim \frac{H^2}{\Delta_\zeta^{1/2}}\ ,\qquad \Lambda^2\lesssim \frac{1}{C^{1/6}}\frac{H^2}{\Delta_\zeta^{2/3}}\left(\frac{\tilde\mu}{H}\right)^{\Delta/6}\ .
\eea
We see that this leaves large room for $\Lambda\gg H$ with the trispectrum being naturally the leading signal. Notice that already imposing the signal from the trispectrum to be detectable, $\tau_{\rm NL}\Delta_\zeta\gtrsim 1$, implies $\tilde\mu^2\sim\Lambda_U^2\lesssim H^2/\Delta_\zeta^{1/\Delta}$. So the limits above represent a mildly stronger constraint.

\section{Details of the Shape Calculation}\label{app:details}

In section Section \ref{sec:shape} we discussed the shape of the bispectrum computed numerically from~(\ref{equ:bispectrum}).  To simplify the numerical calculation, one can perform several integrals analytically before performing the numerical integration. This is important because naive numerical integration leads to a 6-dimensional integral that is hard to evaluate.  In this appendix we will explain which integrations were performed analytically.

First we need to compute the Fourier transform of the three point function at different times and different $k$'s. Let us work in Minkowski space first. We know that to write it in de Sitter space we simply need to multiply by the relevant conformal factors at the end. Let us start by going to 4d Fourier space:

\bea
\langle\cO_{\k_1}(\tau_1)\cO_{\k_2}(\tau_2)\cO_{\k_3}(\tau_3)\rangle=\int {d\omega_1\over 2\pi}  {d\omega_2 \over 2\pi} { d\omega_3\over 2\pi} \; e^{-i\sum_i\omega_i\tau_i }\;\langle\cO_{k_1^\mu}\cO_{k_2^\mu}\cO_{k_3^\mu}\rangle\ ,
\eea
where all correlation functions are meant to be anti-time-ordered.

We know how to write the 3-point function in real space. So, let us do a 4d Fourier transform:
\bea
=\int {d\omega_1\over 2\pi}  {d\omega_2 \over 2\pi} { d\omega_3\over 2\pi}\; e^{-i \sum_i\omega_i\tau_i } \ \int d^4 x_1 d^4 x_2 d^4 x_3\; e^{-i \sum k_i^\mu x_{i,\mu}} C\frac{1}{\left(x_{12}^2\right)^{\Delta/2}}\frac{1}{\left(x_{23}^2\right)^{\Delta/2}}\frac{1}{\left(x_{31}^2\right)^{\Delta/2}}\ .
\eea
We know how to do the Fourier transform of the 2-point function $1/(x^2)^{\Delta/2}$, see eq~(\ref{eq:real-2-point}), so let us write it in this way. It looks like we are adding a lot of integrals, but, as we will see, many of them can be done analytically.  We obtain
\bea
&&=C \int {d\omega_1\over 2\pi}  {d\omega_2 \over 2\pi} { d\omega_3\over 2\pi}\; e^{-i\sum_i\omega_i\tau_i } \ \int d^4 x_1 d^4 x_2 d^4 x_3\; e^{-i \sum k_i^\mu x_{i,\mu}}\;\left(\frac{(2\pi)^2}{4^{\tfrac{\Delta}{2}-1}}\frac{\Gamma(2-\tfrac{\Delta}{2})}{\Gamma(\tfrac{\Delta}{2})}\right)^3\times\nonumber\\ 
&&\qquad \int {d^4p_1\over (2\pi)^4}\,{d^4p_2\over (2\pi)^4}\,{d^4p_3\over (2\pi)^4}\; e^{-i\left[ p_1^\mu x_{12,\mu}+ p_2^\mu x_{23,\mu}+ p_3^\mu x_{31,\mu}\right]} \left(p_1^2\right)^{\Delta/2-2}\left(p_2^2\right)^{\Delta/2-2}\left(p_3^2\right)^{\Delta/2-2}\ .
\eea
The $x_i$-integrals lead to three $\delta$-four functions of the form
\beq
 (2\pi)^4\delta^{(4)}(\sum k^\mu_i)\; (2\pi)^4\delta^{(4)}(p^\mu_1-(p^\mu_3+k^\mu_1))\; (2\pi)^4\delta^{(4)}(p^\mu_2-(p^\mu_3-k^\mu_3))\ .
\eeq
Notice that
\beq
 (2\pi)^4\delta^{(4)}(\sum k^\mu_i)= (2\pi)^3\delta^{(3)}(\sum_i \k_i)\  (2\pi)\delta(\sum_i \omega_i)\ .
\eeq
The first term is the usual 3-delta function of spatial-momentum conservation. We can drop it by adding a $'$ to our correlation function. Now the integral in $p_1$ and $p_2$ can be done saturating the $\delta$-function, and we are left with
\bea
&&=C \int {d\omega_1\over 2\pi}  {d\omega_2 \over 2\pi} { d\omega_3\over 2\pi}\; (2\pi) \delta\left(\sum_i \omega_i\right)\; e^{-i\sum_i\omega_i\tau_i }\ \left(\frac{(2\pi)^2}{4^{\tfrac{\Delta}{2}-1}}\frac{\Gamma(2-\tfrac{\Delta}{2})}{\Gamma(\tfrac{\Delta}{2})}\right)^3\times \\ \nonumber
&&\qquad \int {d^4p_3\over  (2\pi)^4}  \left((p_3+k_1)^2\right)^{\Delta/2-2}\left((p_3-k_3)^2\right)^{\Delta/2-2}\left(p_3^2\right)^{\Delta/2-2}\ .
\eea
The integral in $d^4p_3$ can be done after inserting two Feynman parameters. Let us use that (see for example  App.~F of~\cite{Antoniadis:2011ib}):
\bea
&&A^{-\alpha} B^{-\beta} C^{-\gamma}=\frac{\Gamma(\alpha+\beta+\gamma)}{\Gamma(\alpha)\Gamma(\beta)\Gamma(\gamma)}\int_0^1du\int_0^1 dv \times\\  \nonumber
&& u^{\alpha-1} (1-u)^{\beta-1} v^{\alpha+\beta-1}(1-v)^{\gamma-1}\left[u v A+(1-u) v B+(1-v) C\right]^{-\alpha-\beta-\gamma}\ .
\eea
We then get:
\bea
&&=C \int {d\omega_1\over 2\pi}  {d\omega_2 \over 2\pi} { d\omega_3\over 2\pi}\; (2\pi)\delta(\sum \omega_i) e^{-\sum_i\omega_i\tau_i } \left(\frac{(2\pi)^2}{4^{\tfrac{\Delta}{2}-1}}\frac{\Gamma(2-\Delta/2)}{\Gamma(\Delta/2)}\right)^3\frac{\Gamma(6-\tfrac{3\Delta}{2})}{\Gamma(2-\tfrac{\Delta}{2})^3}\times\\ \nonumber
&&\int_0^1 du\int_0^1dv\;  \left(u(1-v)(1-u)\right)^{1-\Delta/2} v^{3-\Delta}\int {d^4p_3\over  (2\pi)^4} \left(\tilde p_3^2+M^2\right)^{ \tfrac{3 \Delta}{2}-6}\ ,
\eea
where 
\bea
&&M^2(u,v,\{k_i\})=\left(u v (1-v) k_1^2+u v^2 (1-u) k_2^2+v (1-u)(1-v)k_3^2\right), \quad k_i^2=k_i^\mu k_{i,\mu}=\omega_i^2+\vec k_i^2,\nonumber\\ 
\eea
and where we have shifted the variable of integration $p_3$ to complete the square and used that 
\beq
\sum k_i^\mu=0\quad\Rightarrow\quad k_1\cdot k_2=\frac{k_3^2-k_1^2-k_2^2}{2}\ .
\eeq
The $\tilde p_3$ integrals is equal to
\beq
\int \frac{d^4\tilde p_3}{(2\pi)^4}\; \left(\tilde p_3^2+M^2\right)^{\tfrac{3 \Delta}{2}-6} =  \frac{\Gamma(4-\tfrac{3\Delta}{2} )}{ (4 \pi)^2 \Gamma(6-\tfrac{3\Delta}{2} )} \left(M^2\right){}^{\tfrac{3\Delta}{2} -4}\ .
\eeq
We obtain:
\bea \nonumber
&&\langle\cO_{\k_1}(\tau_1)\cO_{\k_2}(\tau_2)\cO_{\k_3}(\tau_3)\rangle'=C \int {d\omega_1\over 2\pi}  {d\omega_2 \over 2\pi} { d\omega_3\over 2\pi}\; (2\pi) \delta\left(\sum \omega_i\right) e^{-i \sum_i\omega_i\tau_i } \ \left(\frac{(2\pi)^2}{4^{\Delta/2-1}}\frac{\Gamma(2-\Delta/2)}{\Gamma(\Delta/2)}\right)^3\\ 
&&\int_0^1 du\int_0^1 dv\;  \frac{\Gamma(4-\tfrac{3\Delta}{2} )}{  (4 \pi)^2 \Gamma(2-\Delta/2)^3}   \left(u(1-v)(1-u)\right)^{1-\Delta/2} v^{3-\Delta}  \left(M^2(u,v,\{k_i\})\right){}^{\tfrac{3\Delta}{2} -4} \ .
\eea
The $v$ integral can be done analytically (by Mathematica, not by us!), finally obtaining:
\bea \nonumber
&&\langle\cO_{\k_1}(\tau_1)\cO_{\k_2}(\tau_2)\cO_{\k_3}(\tau_3)\rangle'=-C\frac{2^{(2-\Delta)} (2\pi)^6 (\Delta-2) }{\Gamma \left(2- \Delta/2\right)^3 \Gamma \left(\frac{\Delta}{2}\right)^3} \csc \left(\pi  \Delta/2\right) \Gamma
   \left(4-3 \Delta/2\right) \Gamma \left(2-\Delta/2\right)^3\;\times \nonumber\\ \nonumber 
   && \int_0^1 du  \int {d\omega_1\over 2\pi}  {d\omega_2 \over 2\pi} { d\omega_3\over 2\pi}\; (2\pi) \delta\left(\sum \omega_i\right) e^{-i \sum_i\omega_i\tau_i }\; \times\\ \nonumber
   && \quad\ ((1-u) u)^{1-\frac{\Delta}{2}}\left(u \left(k_1^2+\omega_1^2-\omega_3^2\right)+k_3^2 (-(u-1))+\omega_3^2\right)^{\frac{3 \Delta}{2}-4}
    \,\times \\ 
&&    _2F_1\left(4-\frac{3
   \Delta}{2},\frac{\Delta}{2};2;1-\frac{(u-1) u \left(k_2^2+\omega_2^2\right)}{(u-1)
   k_3^2+(u-1) \omega_3^2-u \left(k_1^2+\omega_1^2\right)}\right)\ .
   \eea

Now we are ready to deal with our expression for the bispectrum (\ref{eq:bispectrum}). We notice that the $\tau_{1,2,3}$ integrals can be done analytically. Then we can do analytically also the integral over one of the $\omega$'s, as it is just given by the saturation of the delta function. We do not give the result here directly, as it is not illuminating. It is just better to ask Mathematica to do it. At this point we are left with three integrals to do: one over $u$, and two over the remaining two $\omega$'s. By rescaling the variables, it is easy to see that the bispectrum scales as $k^{-6}$, signalling its scale invariance. We can do the resulting integrals only numerically, and Mathematica does them in a few seconds.

\newpage
\begingroup\raggedright

\endgroup


\begin{thebibliography}{10}

\baselineskip=14.5pt

\bibitem{Senatore:2010wk}
  L.~Senatore and M.~Zaldarriaga,
  ``The Effective Field Theory of Multifield Inflation,''
  JHEP {\bf 1204} (2012) 024
  [arXiv:1009.2093 [hep-th]].



\bibitem{LopezNacir:2011kk} 
  D.~Lopez Nacir, R.~A.~Porto, L.~Senatore and M.~Zaldarriaga,
  ``Dissipative effects in the Effective Field Theory of Inflation,''
  JHEP {\bf 1201}, 075 (2012)
  [arXiv:1109.4192 [hep-th]].


\bibitem{Juan}
J.~M.~Maldacena,
  ``Non-Gaussian features of primordial fluctuations in single field inflationary models,''
  JHEP {\bf 0305}, 013 (2003)
  [astro-ph/0210603].

\bibitem{fnLmulti}

D.~S.~Salopek and J.~R.~Bond,
  ``Nonlinear evolution of long wavelength metric fluctuations in inflationary models,''
  Phys.\ Rev.\ D {\bf 42}, 3936 (1990).

A.~D.~Linde and V.~F.~Mukhanov,
  ``Nongaussian isocurvature perturbations from inflation,''
  Phys.\ Rev.\ D {\bf 56}, 535 (1997)
  [astro-ph/9610219].

G.~Dvali, A.~Gruzinov and M.~Zaldarriaga,
  ``Cosmological perturbations from inhomogeneous reheating, freezeout, and mass domination,''
  Phys.\ Rev.\ D {\bf 69}, 083505 (2004)
  [astro-ph/0305548].

 L.~Kofman,
  ``Probing string theory with modulated cosmological fluctuations,''
  astro-ph/0303614.

 M.~Zaldarriaga,
  ``Non-Gaussianities in models with a varying inflaton decay rate,''
  Phys.\ Rev.\ D {\bf 69}, 043508 (2004)
  [astro-ph/0306006].

  
\bibitem{QSFI}
  X.~Chen and Y.~Wang,
  ``Quasi-Single Field Inflation and Non-Gaussianities,''
  JCAP {\bf 1004}, 027 (2010)
  [arXiv:0911.3380 [hep-th]].

  X.~Chen and Y.~Wang,
  ``Large non-Gaussianities with Intermediate Shapes from Quasi-Single Field Inflation,''
  Phys.\ Rev.\ D {\bf 81}, 063511 (2010)
  [arXiv:0909.0496 [astro-ph.CO]].
 
  E.~Sefusatti, J.~R.~Fergusson, X.~Chen and E.~P.~S.~Shellard,
  ``Effects and Detectability of Quasi-Single Field Inflation in the Large-Scale Structure and Cosmic Microwave Background,''
  JCAP {\bf 1208}, 033 (2012)
  [arXiv:1204.6318 [astro-ph.CO]].

\bibitem{Dans} 
  D.~Baumann and D.~Green,
  ``Signatures of Supersymmetry from the Early Universe,''
  Phys.\ Rev.\ D {\bf 85}, 103520 (2012)
  [arXiv:1109.0292 [hep-th]].






\bibitem{ubounds} 
  X.~Dong, B.~Horn, E.~Silverstein and G.~Torroba,
  ``Unitarity bounds and RG flows in time dependent quantum field theory,''  Phys.\ Rev.\ D {\bf 86}, 025013 (2012)  [arXiv:1203.1680 [hep-th]].  







\bibitem{inflthroats}

 D.~Baumann and L.~McAllister,
  ``Advances in Inflation in String Theory,''
  Ann.\ Rev.\ Nucl.\ Part.\ Sci.\  {\bf 59}, 67 (2009)
  [arXiv:0901.0265 [hep-th]].

L.~McAllister and E.~Silverstein,
  ``String Cosmology: A Review,''
  Gen.\ Rel.\ Grav.\  {\bf 40}, 565 (2008)
  [arXiv:0710.2951 [hep-th]].

\bibitem{GKPKKLT}
L.~Randall and R.~Sundrum,
  ``A Large mass hierarchy from a small extra dimension,''
  Phys.\ Rev.\ Lett.\  {\bf 83}, 3370 (1999)
  [hep-ph/9905221].

S.~B.~Giddings, S.~Kachru and J.~Polchinski,
  ``Hierarchies from fluxes in string compactifications,''
  Phys.\ Rev.\ D {\bf 66}, 106006 (2002)
  [hep-th/0105097].

S.~Kachru, R.~Kallosh, A.~D.~Linde and S.~P.~Trivedi,
  ``De Sitter vacua in string theory,''
  Phys.\ Rev.\ D {\bf 68}, 046005 (2003)
  [hep-th/0301240].

\bibitem{tunneling}

S.~Dimopoulos, S.~Kachru, N.~Kaloper, A.~E.~Lawrence and E.~Silverstein,
  ``Generating small numbers by tunneling in multithroat compactifications,''
  Int.\ J.\ Mod.\ Phys.\ A {\bf 19}, 2657 (2004)
  [hep-th/0106128].
  
S.~Dimopoulos, S.~Kachru, N.~Kaloper, A.~E.~Lawrence and E.~Silverstein,
  ``Small numbers from tunneling between brane throats,''
  Phys.\ Rev.\ D {\bf 64}, 121702 (2001)
  [hep-th/0104239].


\bibitem{QCDmonodromy} 


  D.~Baumann and D.~Green,
  ``Desensitizing Inflation from the Planck Scale,''
  JHEP {\bf 1009}, 057 (2010)
  [arXiv:1004.3801 [hep-th]].

  D.~Baumann and D.~Green,
  ``Inflating with Baryons,''
  JHEP {\bf 1104}, 071 (2011)
  [arXiv:1009.3032 [hep-th]].

A.~Maleknejad and M.~M.~Sheikh-Jabbari,
  ``Gauge-flation: Inflation From Non-Abelian Gauge Fields,''
  arXiv:1102.1513 [hep-ph].
  
  S.~Dubovsky, A.~Lawrence and M.~M.~Roberts,
  ``Axion monodromy in a model of holographic gluodynamics,''
  JHEP {\bf 1202}, 053 (2012)
  [arXiv:1105.3740 [hep-th]].
  
 P.~Adshead and M.~Wyman,
  ``Chromo-Natural Inflation: Natural inflation on a steep potential with classical non-Abelian gauge fields,''
  Phys.\ Rev.\ Lett.\  {\bf 108}, 261302 (2012)
  [arXiv:1202.2366 [hep-th]].

\bibitem{DBI}
E.~Silverstein and D.~Tong,
  ``Scalar speed limits and cosmology: Acceleration from D-cceleration,''
  Phys.\ Rev.\ D {\bf 70}, 103505 (2004)
  [hep-th/0310221];

M.~Alishahiha, E.~Silverstein and D.~Tong,
  ``DBI in the sky,''
  Phys.\ Rev.\ D {\bf 70}, 123505 (2004)
  [hep-th/0404084].

\bibitem{generalsingle} 
  X.~Chen, M.~-x.~Huang, S.~Kachru and G.~Shiu,
  ``Observational signatures and non-Gaussianities of general single field inflation,''
  JCAP {\bf 0701}, 002 (2007)
  [hep-th/0605045].

\bibitem{EFT} 
  C.~Cheung, P.~Creminelli, A.~L.~Fitzpatrick, J.~Kaplan and L.~Senatore,
  ``The Effective Field Theory of Inflation,''
  JHEP {\bf 0803}, 014 (2008)
  [arXiv:0709.0293 [hep-th]].

\bibitem{Senatore:2009gt}
  L.~Senatore, K.~M.~Smith and M.~Zaldarriaga,
  ``Non-Gaussianities in Single Field Inflation and their Optimal Limits from the WMAP 5-year Data,''
  JCAP {\bf 1001} (2010) 028
  [arXiv:0905.3746 [astro-ph.CO]].

\bibitem{turning}
  A.~Tolley and M.~Wyman,
  ``The Gelaton Scenario: Equilateral Non-Gaussianity from Multi-Field Dynamics,''
  Phys.\ Rev.\  {\bf D81}, 043502 (2010).

  S.~Cremonini, Z.~Lalak, and K.~Turzynski,
  ``Strongly Coupled Perturbations in Two-Field Inflationary Models,''
  [arXiv:1010.3021 [hep-th]].

  A.~Achucarro {\it et al.},
  ``Features of Heavy Physics in the CMB Power Spectrum,''
  [arXiv:1010.3693 [hep-ph]].
  
  D.~Baumann and D.~Green,
  ``Equilateral Non-Gaussianity and New Physics on the Horizon,''
  JCAP {\bf 1109}, 014 (2011)
  [arXiv:1102.5343 [hep-th]].



\bibitem{Weinberg:2005vy} 
  S.~Weinberg,
  ``Quantum contributions to cosmological correlations,''
  Phys.\ Rev.\ D {\bf 72}, 043514 (2005)
  [hep-th/0506236].

\bibitem{Behbahani:2012be} 
  S.~R.~Behbahani and D.~Green,
  ``Collective Symmetry Breaking and Resonant Non-Gaussianity,''
  arXiv:1207.2779 [hep-th].


\bibitem{Senatore:2009cf}
  L.~Senatore and M.~Zaldarriaga,
  ``On Loops in Inflation,''
  JHEP {\bf 1012} (2010) 008
  [arXiv:0912.2734 [hep-th]].
  G.~L.~Pimentel, L.~Senatore and M.~Zaldarriaga,
  ``On Loops in Inflation III: Time Independence of zeta in Single Clock Inflation,''
  JHEP {\bf 1207} (2012) 166
  [arXiv:1203.6651 [hep-th]].

\bibitem{Dalal:2007cu}
  N.~Dalal, O.~Dore, D.~Huterer and A.~Shirokov,
  ``The imprints of primordial non-gaussianities on large-scale structure: scale dependent bias and abundance of virialized objects,''
  Phys.\ Rev.\ D {\bf 77} (2008) 123514
  [arXiv:0710.4560 [astro-ph]].

\bibitem{Baldauf:2011bh}
  T.~Baldauf, U.~Seljak, L.~Senatore and M.~Zaldarriaga,
  ``Galaxy Bias and non-Linear Structure Formation in General Relativity,''
  JCAP {\bf 1110} (2011) 031
  [arXiv:1106.5507 [astro-ph.CO]].

\bibitem{Slosar:2008hx}
  A.~Slosar, C.~Hirata, U.~Seljak, S.~Ho and N.~Padmanabhan,
  ``Constraints on local primordial non-Gaussianity from large scale structure,''
  JCAP {\bf 0808} (2008) 031
  [arXiv:0805.3580 [astro-ph]].


\bibitem{Baldauf:2010vn}
  T.~Baldauf, U.~Seljak and L.~Senatore,
  ``Primordial non-Gaussianity in the Bispectrum of the Halo Density Field,''
  JCAP {\bf 1104} (2011) 006
  [arXiv:1011.1513 [astro-ph.CO]].


\bibitem{Norena:2012yi}
  J.~Norena, L.~Verde, G.~Barenboim and C.~Bosch,
  ``Prospects for constraining the shape of non-Gaussianity with the scale-dependent bias,''
  JCAP {\bf 1208} (2012) 019
  [arXiv:1204.6324 [astro-ph.CO]].


\bibitem{Sefusatti:2012ye}
  E.~Sefusatti, J.~R.~Fergusson, X.~Chen and E.~P.~S.~Shellard,
  ``Effects and Detectability of Quasi-Single Field Inflation in the Large-Scale Structure and Cosmic Microwave Background,''
  JCAP {\bf 1208} (2012) 033
  [arXiv:1204.6318 [astro-ph.CO]].

\bibitem{Seljak:2008xr}
  U.~Seljak,
  ``Extracting primordial non-gaussianity without cosmic variance,''
  Phys.\ Rev.\ Lett.\  {\bf 102} (2009) 021302
  [arXiv:0807.1770 [astro-ph]].



\bibitem{Babich:2004gb} 
  D.~Babich, P.~Creminelli and M.~Zaldarriaga,
  ``The Shape of non-Gaussianities,''
  JCAP {\bf 0408}, 009 (2004)
  [astro-ph/0405356].


\bibitem{Creminelli:2005hu}
  P.~Creminelli, A.~Nicolis, L.~Senatore, M.~Tegmark and M.~Zaldarriaga,
  ``Limits on non-gaussianities from wmap data,''
  JCAP {\bf 0605} (2006) 004
  [astro-ph/0509029].
  
  

\bibitem{Bennett:2012fp}
  C.~L.~Bennett, D.~Larson, J.~L.~Weiland, N.~Jarosik, G.~Hinshaw, N.~Odegard, K.~M.~Smith and R.~S.~Hill {\it et al.},
  ``Nine-Year Wilkinson Microwave Anisotropy Probe (WMAP) Observations: Final Maps and Results,''
  arXiv:1212.5225 [astro-ph.CO].



\bibitem{Caracciolo:2009bx}
  F.~Caracciolo and V.~S.~Rychkov,
  ``Rigorous Limits on the Interaction Strength in Quantum Field Theory,''
  Phys.\ Rev.\ D {\bf 81} (2010) 085037
  [arXiv:0912.2726 [hep-th]].
  

\bibitem{Poland:2011ey}
  D.~Poland, D.~Simmons-Duffin and A.~Vichi,
  ``Carving Out the Space of 4D CFTs,''
  JHEP {\bf 1205} (2012) 110
  [arXiv:1109.5176 [hep-th]].



\bibitem{SY} 
  T.~Suyama and M.~Yamaguchi,
  ``Non-Gaussianity in the Modulated Reheating Scenario,''
  Phys.\ Rev.\ D {\bf 77}, 023505 (2008).
  
\bibitem{Sugiyama:2011jt} 
  N.~Sugiyama, E.~Komatsu, and T.~Futamase,
  ``Non-Gaussianity Consistency Relation for Multi-Field Inflation,''
  Phys.\ Rev.\ Lett.\  {\bf 106}, 251301 (2011).

\bibitem{Lewis:2011au} 
  A.~Lewis,
  `The Real Shape of Non-Gaussianities,''
  JCAP {\bf 1110}, 026 (2011).

\bibitem{Smith:2011if} 
  K.~Smith, M.~LoVerde, and M.~Zaldarriaga,
  ``A Universal Bound on $N$-point Correlations from Inflation,''
  Phys.\ Rev.\ Lett.\  {\bf 107}, 191301 (2011).
  
  \bibitem{Quasi4}
V.~Assassi, D.~Baumann, and D.~Green,
  ``On Soft Limits of Inflationary Correlation Functions,''
  arXiv:1204.4207 [hep-th].
  
\bibitem{Kehagias:2012pd} 
  A.~Kehagias and A.~Riotto,
  ``Operator Product Expansion of Inflationary Correlators and Conformal Symmetry of de Sitter,''
  arXiv:1205.1523 [hep-th].


\bibitem{Smith:2010gx} 
  K.~M.~Smith and M.~LoVerde,
  ``Local stochastic non-Gaussianity and N-body simulations,''
  JCAP {\bf 1111}, 009 (2011)
  [arXiv:1010.0055 [astro-ph.CO]].

\bibitem{Baumann:2012bc} 
  D.~Baumann, S.~Ferraro, D.~Green and K.~M.~Smith,
  ``Stochastic Bias from Non-Gaussian Initial Conditions,''
  arXiv:1209.2173 [astro-ph.CO].

\bibitem{IntriligatorGrinstein}
 B.~Grinstein, K.~A.~Intriligator and I.~Z.~Rothstein,
  ``Comments on Unparticles,''
  Phys.\ Lett.\ B {\bf 662}, 367 (2008)
  [arXiv:0801.1140 [hep-ph]].

\bibitem{nicolis} 
  S.~Endlich, A.~Nicolis and J.~Wang,
  ``Solid Inflation,''
  arXiv:1210.0569 [hep-th].


\bibitem{Senatore:2010jy}
  L.~Senatore and M.~Zaldarriaga,
  ``A Naturally Large Four-Point Function in Single Field Inflation,''
  JCAP {\bf 1101} (2011) 003
  [arXiv:1004.1201 [hep-th]].


\bibitem{Antoniadis:2011ib} 
  I.~Antoniadis, P.~O.~Mazur and E.~Mottola,
  ``Conformal Invariance, Dark Energy, and CMB Non-Gaussianity,''
  JCAP {\bf 1209}, 024 (2012)
  [arXiv:1103.4164 [gr-qc]].

\end{thebibliography}
\end{document}